\documentclass{emulateapj}

\slugcomment{Not to appear in Nonlearned J., 45.}

\shorttitle{Dense molecular gas in two ULIRGs at $z \sim 1.7$}
\shortauthors{Oteo et al.}

\usepackage{natbib}
\usepackage{CJK}

\begin{document}

\title{High dense gas fraction in intensely star-forming dusty galaxies at high redshift}

\author{
I. Oteo\altaffilmark{1,2}, 
Z-Y. Zhang\altaffilmark{1,2}, 
C. Yang\altaffilmark{3,4,5,6,7}, 
R. J. Ivison\altaffilmark{2,1}, 
A. Omont\altaffilmark{6,7},
M. Bremer\altaffilmark{8}, 
S. Bussmann\altaffilmark{2,1},
A. Cooray\altaffilmark{10}, 
P. Cox\altaffilmark{11}, 
H. Dannerbauer\altaffilmark{12,13}, 
L. Dunne\altaffilmark{1,14}, 
S. Eales\altaffilmark{14}
C. Furlanetto\altaffilmark{15}, 
R. Gavazzi\altaffilmark{16},
H. Nayyeri\altaffilmark{10}, 
M. Negrello\altaffilmark{14}, 
R. Neri\altaffilmark{18}, 
D. Riechers\altaffilmark{9}, 
P. Van der Werf\altaffilmark{19}
}
\affil{
$^{1}$Institute for Astronomy, University of Edinburgh, Royal Observatory, Blackford Hill, Edinburgh EH9 3HJ \\
$^{2}$European Southern Observatory, Karl-Schwarzschild-Str. 2, 85748 Garching, Germany \\
$^{3}$Purple Mountain Observatory/Key Lab of Radio Astronomy, Chinese Academy of Sciences, Nanjing 210008, PR China \\
$^{4}$Institut d'Astrophysique Spatiale, CNRS, Univ. Paris-Sud, Universit\'e Paris-Saclay, B\^at. 121, 91405 Orsay cedex, France\\
$^{5}$Graduate University of the Chinese Academy of Sciences, 19A Yuquan Road, Shijingshan District, 10049, Beijing, PR China \\
$^{6}$CNRS, UMR 7095, Institut d'Astrophysique de Paris, F-75014, Paris, France\\
$^{7}$UPMC Univ. Paris 06, UMR 7095, Institut d'Astrophysique de Paris, F-75014, Paris, France\\
$^{8}$H.H. Wills Physics Laboratory, University of Bristol, Tyndall Avenue, Bristol BS8 1TL, UK\\
$^9$Department of Astronomy, Space Science Building, Cornell University, Ithaca, NY, 14853-6801\\
$^{10}$Department of Physics and Astronomy, University of California, Irvine, CA 92697\\
$^{11}$Joint ALMA Observatory - ESO, Av. Alonso de Cordova, 3104, Santiago, Chile\\
$^{12}$Universidad de La Laguna, Dpto. Astrof\'isica, E-38206 La Laguna, Tenerife, Spain\\
$^{13}$Instituto de Astrof\'isica de Canarias, E-38205 La Laguna, Tenerife, Spain\\
$^{14}$School of Physics and Astronomy, Cardiff University, The Parade, Cardiff CF24 3AA, UK\\
$^{15}$School of Physics and Astronomy, Nottingham University, University Park, Nottingham, NG7 2RD, UK\\
$^{16}$Institut d'Astrophysique de Paris, UMR7095 CNRS-Universit\'e Pierre et Marie Curie, 98bis bd Arago, F-75014 Paris, France\\
$^{17}$BAER Institute/NASA Ames Research Center, Moffet Field, CA 94035, USA\\
$^{18}$IRAM, 300 rue de la piscine, F-38406 Saint-Martin d'Heres, France\\
$^{19}$Leiden Observatory, Leiden University, P.O. Box 9513, NL-2300 RA Leiden, The Netherlands\\
}

\email{ivanoteogomez@gmail.com}

\begin{abstract}

We present ALMA and VLA detections of the dense molecular gas tracers HCN, HCO$^+$ and HNC in two lensed, high-redshift starbursts selected from the {\it Herschel}-ATLAS survey: {\it H}-ATLAS\,J090740.0$-$004200 (SDP.9, $z \sim 1.6$) and {\it H}-ATLAS\,J091043.1$-$000321 (SDP.11, $z \sim 1.8$). ALMA observed the $J = 3-2$ transitions in both sources, while the VLA observed the $J = 1-0$ transitions in SDP.9. We have detected all observed HCN and HCO$^+$ lines in SDP.9 and SDP.11, and also HNC(3--2) in SDP.9. The amplification factors for both galaxies have been determined from sub-arcsec resolution CO and dust emission observations carried out with NOEMA and the SMA. The HNC(1--0)/HCN(1--0) line ratio in SDP.9 suggests the presence of photon-dominated regions, as it happens to most local (U)LIRGs. The CO, HCN and HCO$^+$ SLEDs of SDP.9 are compatible to those found for many local, infrared (IR) bright galaxies, indicating that the molecular gas in local and high-redshift dusty starbursts can have similar excitation conditions. We obtain that the correlation between total IR ($L_{\rm IR}$) and dense line ($L_{\rm dense}$) luminosity in SDP.9 and SDP.11 and local star-forming galaxies can be represented by a single relation. The scatter of the $L_{\rm IR} - L_{\rm dense}$ correlation, together with the lack of sensitive dense molecular gas tracer observations for a homogeneous sample of high-redshift galaxies, prevents us from distinguishing differential trends with redshift. Our results suggest that the intense star formation found in some high-redshift dusty, luminous starbursts is associated with more massive dense molecular gas reservoirs and higher dense molecular gas fractions.

\end{abstract}


\section{Introduction}\label{intro}

Active star formation in galaxies occurs in dense regions within molecular clouds. Studying the dense interstellar medium is, therefore, of fundamental importance for our understanding of the formation and evolution of star-forming galaxies. The critical densities of rotational transitions are proportional to $\mu^2 \, \nu_{J+1,J}^3$, where $\mu$ is the dipole moment and $\nu_{J+1,J}$ is the frequency of the rotational level $J$ of the molecule \citep{Shirley2015PASP..127..299S}. Because the dipole moments of the HCN, HNC and HCO$^+$ molecules are almost $30 \times$ times higher than those for CO, the HCN, HNC and HCO$^+$ transitions trace a gas component whose critical density could be $\sim 100-500$ times denser than the gas traced by CO for the same rotational level. 


Because HCN, HCO$^+$ and HNC can only be excited in very dense regions, they trace the places where star formation takes place. This is why correlations have been found between the HCN and total IR luminosity \citep{Gao2004ApJS..152...63G,Privon2015ApJ...814...39P}, although the relation between SFR and dense gas emission can be also affected by other several physical factors \citep{GarciaBurillo2012A&A...539A...8G,Krips2008ApJ...677..262K,Imanishi2007AJ....134.2366I,Imanishi2009AJ....137.3581I,Davies2012A&A...537A.133D,Chen2016arXiv161200459C}. Actually, there are places with dense gas but without star formation, and there are also places with star formation, but few dense gas. One of the most obvious cases is the Galactic central region, especially the central molecular zone, where the average ${\rm H_2}$ density is $> 10^{4} \, {\rm cm}^{-3}$ and there is few star formation comparing to normal molecular clouds \citep{Longmore2013MNRAS.429..987L,Kauffmann2013ApJ...765L..35K}. It happens the same in some local galaxies, such the antennae galaxy or the nuclear region of NGC\,4039, or NGC\,1068 \citep{Ueda2012ApJ...745...65U,GarciaBurillo2014A&A...567A.125G}. On the other hand, \cite{Papadopoulos2014ApJ...788..153P} showed that Arp\,193 has much less ($< 10\%$) dense gas mass than NGC\,6240, while they have similar total IR luminosities and are not heavily affected by AGN.

One of the main challenges of studying the dense molecular gas via HCN, HCO$^+$ and HNC transitions is that they are typically faint, about an order of magnitude fainter than CO lines in LIRGs and ULIRGs, and up to two order of magnitude for normal galaxies \citep{Gao2004ApJ...606..271G,Gao2004ApJS..152...63G,Bussmann2008ApJ...681L..73B,Juneau2009ApJ...707.1217J}. Therefore, whereas dense molecular gas tracers have been detected in a fairly large number of galaxies in the local Universe \citep{Aalto2002A&A...381..783A,Gao2004ApJ...606..271G,Gao2004ApJS..152...63G,Papadopoulos2007ApJ...656..792P,Baan2008A&A...477..747B,GraciaCarpio2008A&A...479..703G,Bussmann2008ApJ...681L..73B,Juneau2009ApJ...707.1217J,Liu2015ApJ...805...31L,Privon2015ApJ...814...39P}, the detection of HCN, HCO$^+$ and HNC at high redshift has been limited to a handful of extreme sources such as QSOs where the probability of detection was higher \citep{Gao2007ApJ...660L..93G,Riechers2011ApJ...726...50R}. One way to overcome the difficulty of detecting dense molecular gas tracers is taking advantage of the amplification provided by gravitational lensing. The combination of intrinsic FIR brightness, lens amplification and sensitive instrumentation like ALMA and VLA enables to carry out studies of faint emission lines that otherwise would require very long integration times.

\begin{table}
\caption{\label{table_flux_values}Observed properties of SDP.9 and SDP.11 (not corrected from lens magnification).}
\centering
\begin{tabular}{ccc}
& SDP.9 & SDP.11 \\
\hline\hline
RA & 09:07:40.0 & 09:10:43.1 \\
DEC & $-$00:41:59.8 & $-$00:03:22.8\\
$T_{\rm dust}$ [K]\tablenotemark{(a)} & $43 \pm 2$  & $41 \pm 1$\\
$L_{\rm IR} / 10^{13} \, L_\odot$\tablenotemark{(a)}  & $6.4 \pm 0.1$ & $6.6 \pm 0.1$\\
${\rm SFR \,[M_\odot \, yr^{-1}]}$\tablenotemark{(a)} & $\sim 11,500$ & $\sim 11,800$\\
$z_{\rm spec}$\tablenotemark{(b)} & $1.575 \pm 0.005$ & $1.786 \pm 0.005$\\
$\mu_{\rm CO}$ & $5.8 \pm 2.9$ & $5.5 \pm 1.0$\\
$\mu_{\rm dust}$ & $4.5 \pm 1.9$ & $6.1 \pm 1.2$ \\
$\mu_{\rm stars}$\tablenotemark{(c)} & $6.3 \pm 0.3$ & $4.9 \pm 0.9$\\
\hline
Emission line Fluxes \\
\hline
$\rm ^{12}CO$ $J$=1--0 $I_{\rm CO}$ [${\rm Jy} \, {\rm km} \, {\rm s}^{-1}$] 	&	$1.0 \pm 0.2$	\\
$\rm ^{12}CO$ $J$=3--2 $I_{\rm CO}$ [${\rm Jy} \, {\rm km} \, {\rm s}^{-1}$] 	&	$10.1 \pm 0.6$ & --	\\
$\rm ^{12}CO$ $J$=4--3 $I_{\rm CO}$ [${\rm Jy} \, {\rm km} \, {\rm s}^{-1}$] 	&	-- & $8.9 \pm 1.0$	\\
$\rm ^{12}CO$ $J$=5--4 $I_{\rm CO}$ [${\rm Jy} \, {\rm km} \, {\rm s}^{-1}$]\tablenotemark{(d)} 	&	$25 \pm 5$ & $23 \pm 8$	\\
$\rm ^{12}CO$ $J$=6--5 $I_{\rm CO}$ [${\rm Jy} \, {\rm km} \, {\rm s}^{-1}$]\tablenotemark{(d)} 	&	$33 \pm 7 $ & $29 \pm 10$	\\
$\rm ^{12}CO$ $J$=7--6 $I_{\rm CO}$ [${\rm Jy} \, {\rm km} \, {\rm s}^{-1}$]\tablenotemark{(d)} 	&	-- & $18 \pm 14$	\\
HCN(1--0) [${\rm Jy} \, {\rm km} \, {\rm s}^{-1}$]  & $0.17 \pm 0.06$ &  --\\
HCO$^{+}$(1--0) [${\rm Jy} \, {\rm km} \, {\rm s}^{-1}$]  & $0.12 \pm 0.02$ &  --\ \\
HNC(1--0) [${\rm Jy} \, {\rm km} \, {\rm s}^{-1}$]  & $< 0.08$ &  --\ \\
HCN(3--2) [${\rm Jy} \, {\rm km} \, {\rm s}^{-1}$] 	&	$0.66 \pm 0.11$ &	$0.54 \pm 0.08$\\
HCO$^{+}$(3--2) [${\rm Jy} \, {\rm km} \, {\rm s}^{-1}$] 	&	$0.36 \pm 0.14$ &	$0.36 \pm 0.03$\\
HNC(3--2) [${\rm Jy} \, {\rm km} \, {\rm s}^{-1}$] 	&	$0.42 \pm 0.10$ & $< 0.10$\\
\hline
\hline
\tablenotetext{1}{From \cite{Bussmann2013ApJ...779...25B}}
\tablenotetext{2}{From \cite{Lupu2012ApJ...757..135L}}
\tablenotetext{3}{From \cite{Dye2014MNRAS.440.2013D}}
\tablenotetext{4}{From \cite{Lupu2012ApJ...757..135L}}
\end{tabular}
\end{table} 

In this work, we present VLA and ALMA detections of three dense molecular gas tracers, HCN, HCO$^+$ and HNC, in two lensed ULIRGs selected from the {\it H}-ATLAS \citep{Eales2010PASP..122..499E}: {\it H}-ATLAS\,J090740.0$-$004200 (hereafter SDP.9, at $z \sim 1.6$) and {\it H}-ATLAS\,J091043.1$-$000321 (hereafter SDP.11, at $z \sim 1.8$), see \cite{Negrello2010Sci...330..800N,Negrello2014MNRAS.440.1999N}. The properties of the targets can be found in Table \ref{table_flux_values}, including redshifts, coordinates, unlensed total IR luminosity, SFR and dust temperature. All these values have been taken from \cite{Bussmann2013ApJ...779...25B}. The amplification factors have been obtained from CO and dust emission observations at sub-arcsec spatial resolution carried out with NOEMA and the SMA. The structure of the paper is as follows: \S \ref{data_sets} presents the data used in this work. \S \ref{lens_model_section} explains the lens modeling of the studied sources. Next, \S \ref{section_results_and_discussion} presents the molecular line detections and the study of the physical conditions of the gas in our galaxies. Finally, \S \ref{conclu} summarizes the main conclusions of this work.

\begin{figure*}
\centering
\includegraphics[width=0.50\textwidth]{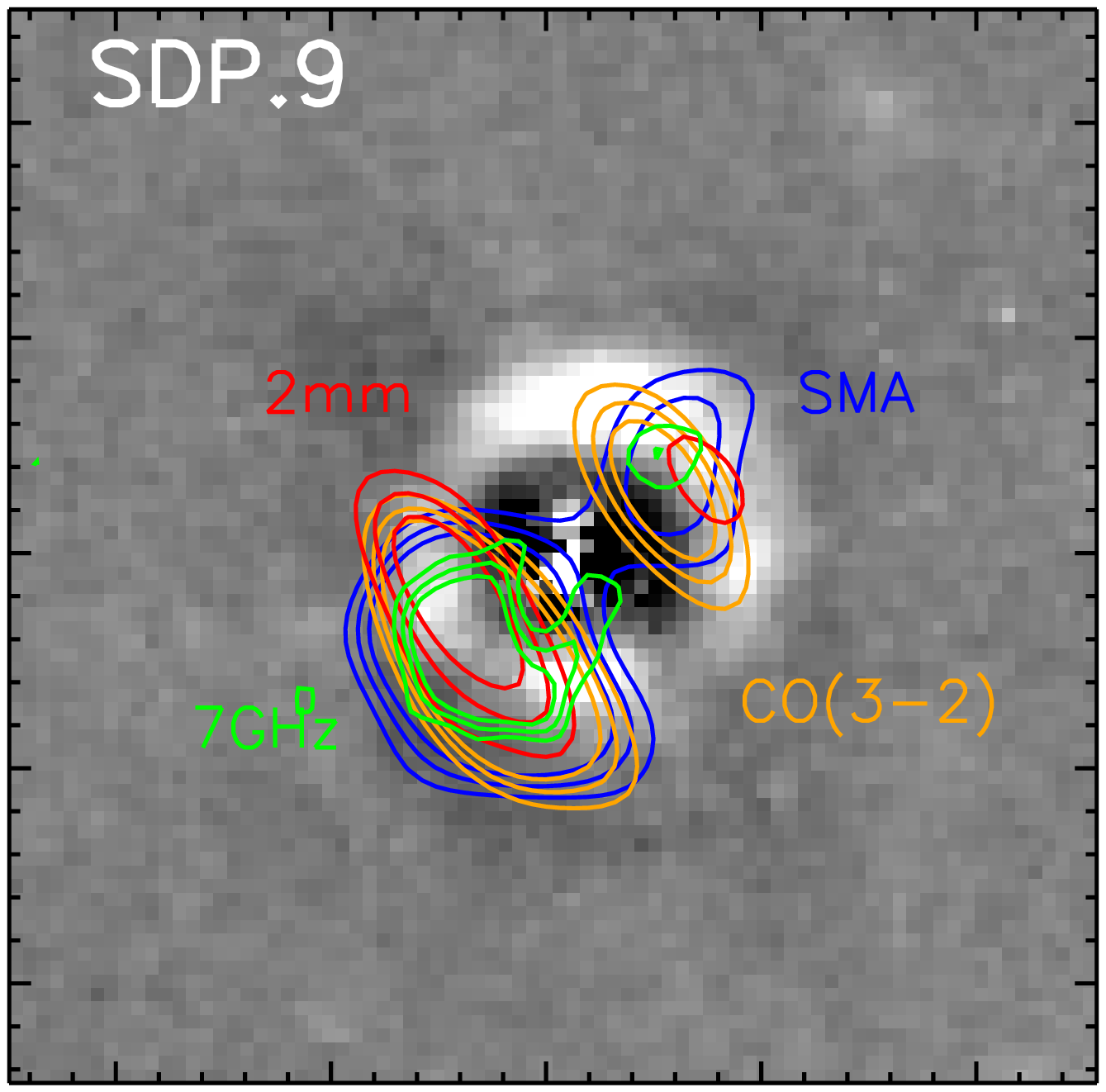}
\hspace{-24mm}
\includegraphics[width=0.50\textwidth]{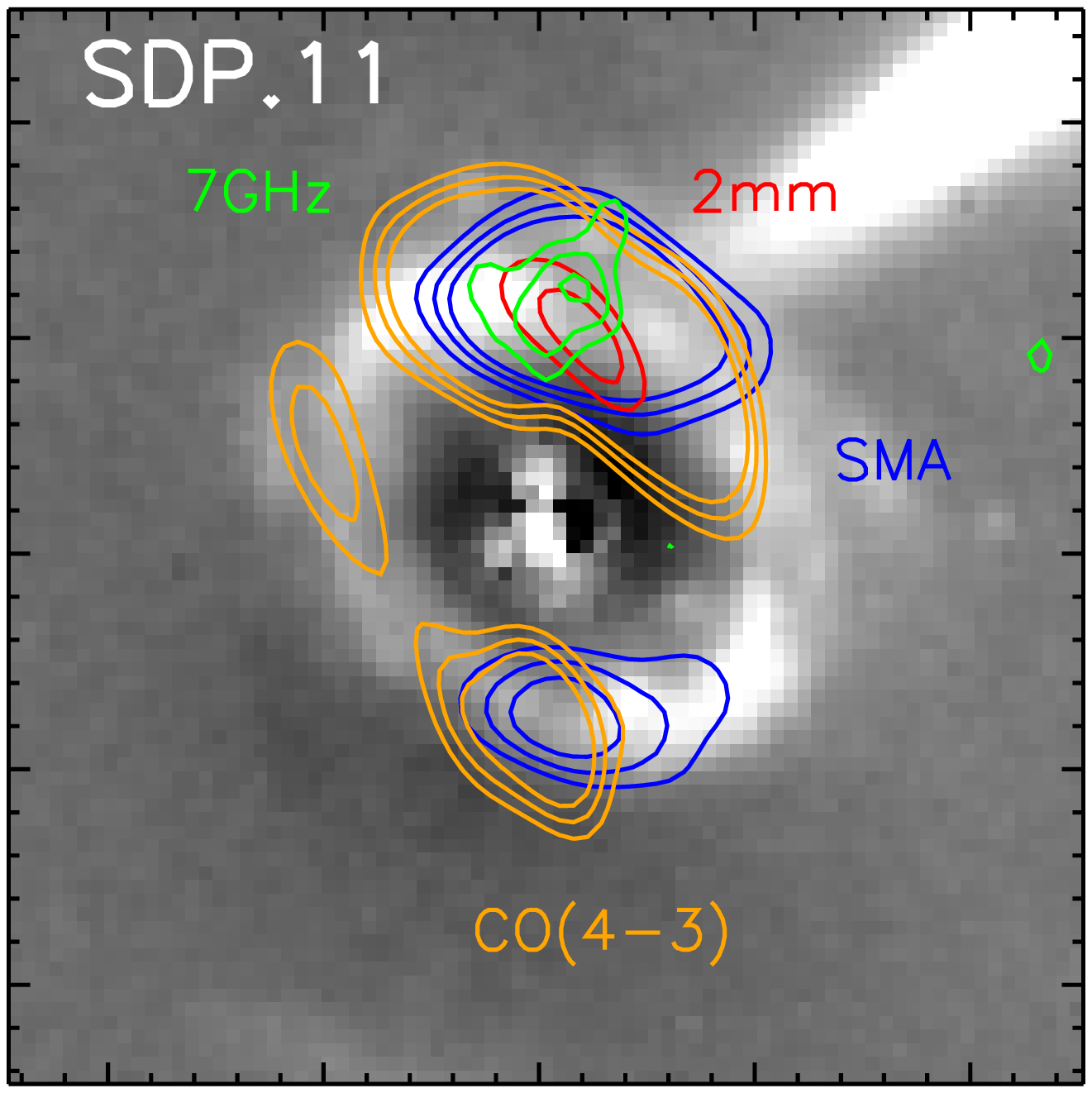}\\
\vspace*{-10mm}
\caption{Resolved stellar, dust, molecular gas and radio continuum emission in the two lensed ULIRGs studied in this work: SDP.9 at $z \sim 1.6$ ({\it left}) and SDP.11 at $z \sim 1.8$ ({\it right}). The background images represent the stellar emission observed by HST (the foreground lenses have been subtracted by fitting their light profiles with GALFIT). The size of each image is $5'' \times 5''$, north is up, east is left. Contours represent the dust (at $870 \, {\rm \mu m}$ --blue-- from the SMA and $2 \, {\rm mm}$ --red-- from PdBI), molecular gas (via the CO(3--2) transition -- orange) and radio continuum emission (from VLA at $7 \, {\rm GHz}$ -- green). All contours are represented from $3\sigma$ to $5\sigma$ in steps of $1\sigma$. The dense molecular gas and $^{12}$CO(1--0) transitions are not shown because they are unresolved due the relatively large beam of the ALMA and VLA observations. 
              }
\label{images_resolved_lenses}
\end{figure*}

\section{Data sets}\label{data_sets}


{\it ALMA observations}:  SDP.9 and SDP.11 were observed with ALMA in the 3\,mm band in compact configuration (project 2012.1.00915.S, PI: R. Lupu) with the aim of detecting their HCN(3-2), HCO$^+$(3-2) and HNC(3-2) emission lines. For each source, the spectral setup of the observations was defined to sample the three lines simultaneously. The data for each galaxy were reduced independently using standard techniques in CASA. Imaging was done using natural weighting to improve the signal to noise of the detections. The ${\rm r.m.s}$ of the cubes is $\sim 0.14 \, {\rm mJy \, beam^{-1}}$ in $100 \, {\rm km \, s^{-1}}$ wide channels, with a restoring beam of $3.7'' \times 2.3''$.

{\it VLA observations}: The National Radio Astronomy Observatory (NRAO) VLA observations used in this paper were carried out on 5th October 2014 when the array was in its DnC-configuration (project 	14B-259, PI: C. Yang) and the spectral setup was defined to cover the HCN(1--0), HCO$^+$(1--0), and HNC(1--0) transitions in SDP.9 simultaneously. The data were manually calibrated by following the standard procedures, including manual flagging and re-calibration. The gain calibrator was J0909+0200, which has a weak emission of $\sim 0.3 \, {\rm Jy}$, we combined all 24 spectral windows (spws) together in the gain calibration to get enough signal to noise, instead of calibrating gain solutions spw by spw. By doing this we assume that the responses of different spws do not vary with frequency. After imaging (also done with natural weighting), the r.m.s reached is ${\rm \sim 0.17 mJy \, beam^{-1}}$ in $17 \, {\rm km \, s^{-1}}$ channels, with a synthesized beam size of $2.37'' \times 0.99''$ at ${\rm PA = 74.7 \, deg}$. No VLA observations targeting the 1--0 transitions of the HCN, HCO$^+$ and HNC emissions are available for SDP.11.

The rest-frame $18\,{\rm GHz}$ emission of SDP.9 and SDP.11 was also observed with the VLA in its most extended configuration at 7 GHz (project 12B-189, PI: R. Ivison). Additionally, the $^{12}$CO(1--0) transition in SDP.9 was also observed with the VLA in C configuration (project 12A-201, PI: R. Ivison). Both datasets were calibrated in {\sc AIPS} following the standard prescriptions, including manual flagging of the RFI emission in C band.


{\it PdBI observations}: The $^{12}$CO(3-2) emission in SDP.9 and SDP.11 was carried out by the PdBI in its most extended configuration during January and February 2011 (programs UAAA and UBAA PIs: P. Cox and R. Ivison). The data were calibrated by using the standard techniques with GILDAS. Imaging was carried out using natural weighting to maximize sensitivity. In SDP.9 we reached an ${\rm r.m.s} \sim 0.8 \, {\rm mJy \, {\rm beam}^{-1}}$ in $50 \, {\rm km \, s^{-1}}$ wide channels, while the noise level in SDP.11 is ${\rm r.m.s} \sim 1.2 \, {\rm mJy \, {\rm beam}^{-1}}$ in $50 \, {\rm km \, s^{-1}}$ wide channels. The median beam size of the spectral cubes are $1.10'' \times 0.46''$ and $0.86'' \times 0.36''$ for SDP.9 and SDP.11, respectively.

All previous datasets are also combined with available dust continuum observations taken with the SMA \citep{Bussmann2013ApJ...779...25B} and stellar emission observed with HST at near-IR wavelengths \citep{Negrello2014MNRAS.440.1999N}. We refer the reader to those works for details on the data calibration and reduction.






\section{Lens modelling}\label{lens_model_section}

We have modeled the lensed dust continuum (observed with the SMA at $880 \, {\rm \mu m}$) and $^{12}$CO(3--2) (observed with the PdBI) emission in SDP.9 and SDP.11 (see Figure \ref{images_resolved_lenses}) by using \verb+uvmcmcfit+ \citep{Bussmann2015arXiv150405256B}, a code designed to model, in the $uv$ plane, the emission of gravitationally lensed galaxies observed with interferometers. Note that we have not modeled the PdBI-observed $2 \, {\rm mm}$ continuum due to the poor signal to noise. We have not modelled any of the emission observed by the VLA and ALMA because the spatial resolution is not enough to resolve the lensed emission. \verb+uvmcmcfit+ assumes a pre-defined shape for the background source (elliptical gaussian profile) and that the lens galaxy can be described by singular isothermal ellipsoid (SIE) profiles. The lensed dust emission in SDP.9 and SDP.11 observed by the SMA was already modeled in \cite{Bussmann2013ApJ...779...25B} with an initial version of \verb+uvmcmcfit+ based on the use of {\sc gravlens} \citep{Keeton2001astro.ph..2340K}. However, for consistency, we have also modeled the SMA emission in our two sources with the last version of \verb+uvmcmcfit+, the same as the one used to model the CO lines. The derived amplification values from dust continuum and CO emission are given in Table \ref{table_flux_values} (uncertainties are derived from the FWHM of the posterior probability distribution function), along with the amplifications derived for the stellar emission by \cite{Dye2014MNRAS.440.2013D} based on the model of the emission observed by HST.


The magnification factors derived from the dust continuum SMA maps will be used to estimate the de-magnified IR luminosities. They will be also used to de-magnify the dense molecular gas emission. This is because the dense molecular gas traces the regions where star formation is taking place, so it might be expected that they are both spatially coincident (but see later a note on differential amplification). The magnification derived from the $^{12}$CO(3--2) emission will be used to derive the de-magnified CO luminosities. Low-$J$ CO emission might be more extended than mid- and high-$J$ CO transitions in dusty high-redshift starbursts \citep{Ivison2011MNRAS.412.1913I,Riechers2011ApJ...739L..31R}, and dust emission might not fully coincident with the dense molecular gas emission. Different spatial distributions of the different ISM components would imply differential amplification \citep{Spilker2015ApJ...811..124S}.  Since there is no high-spatial resolution data for the low-$J$ CO and any of the HCN, HCO$^+$ and HNC transitions, we need to assume that their amplification factors are similar to the derived from the $^{12}$CO(3--2) line and dust continuum emission, respectively. The possible differential magnification between different transitions represents one of the main sources of uncertainty in the analysis presented in this work.

\section{Results and discussion}\label{section_results_and_discussion}


\subsection{Resolved stellar, gas, and dust emission}

Figure \ref{images_resolved_lenses} shows the lensed stellar emission observed with HST (the foreground lensing source has been subtracted using \verb+GALFIT+) compared to the resolved dust (observed by SMA and PdBI), molecular gas (observed by PdBI) and radio continuum (observed by VLA) emission. Dust absorbs the rest-frame UV stellar light emitted by massive OB stars at rest-frame UV, which is, in turn, re-emitted at FIR wavelengths \citep[see for example][]{Lutz2014ARA&A..52..373L}. Therefore, there are two main components contributing to the total SFR in star-forming galaxies, one un-obscured component traced by rest-frame UV emission and one obscured component traced by FIR emission. This simplified picture is exactly what we are witnessing in Figure \ref{images_resolved_lenses}. The dust emission detected by the SMA and PdBI represents the obscured star formation and it is preferentially located in regions where stellar light is faint, because it has been absorbed. On the other hand, stellar emission is detected in dust-poor regions, where no dust or molecular gas emission is detected.

\begin{figure}[!t]
\centering
\includegraphics[width=0.45\textwidth]{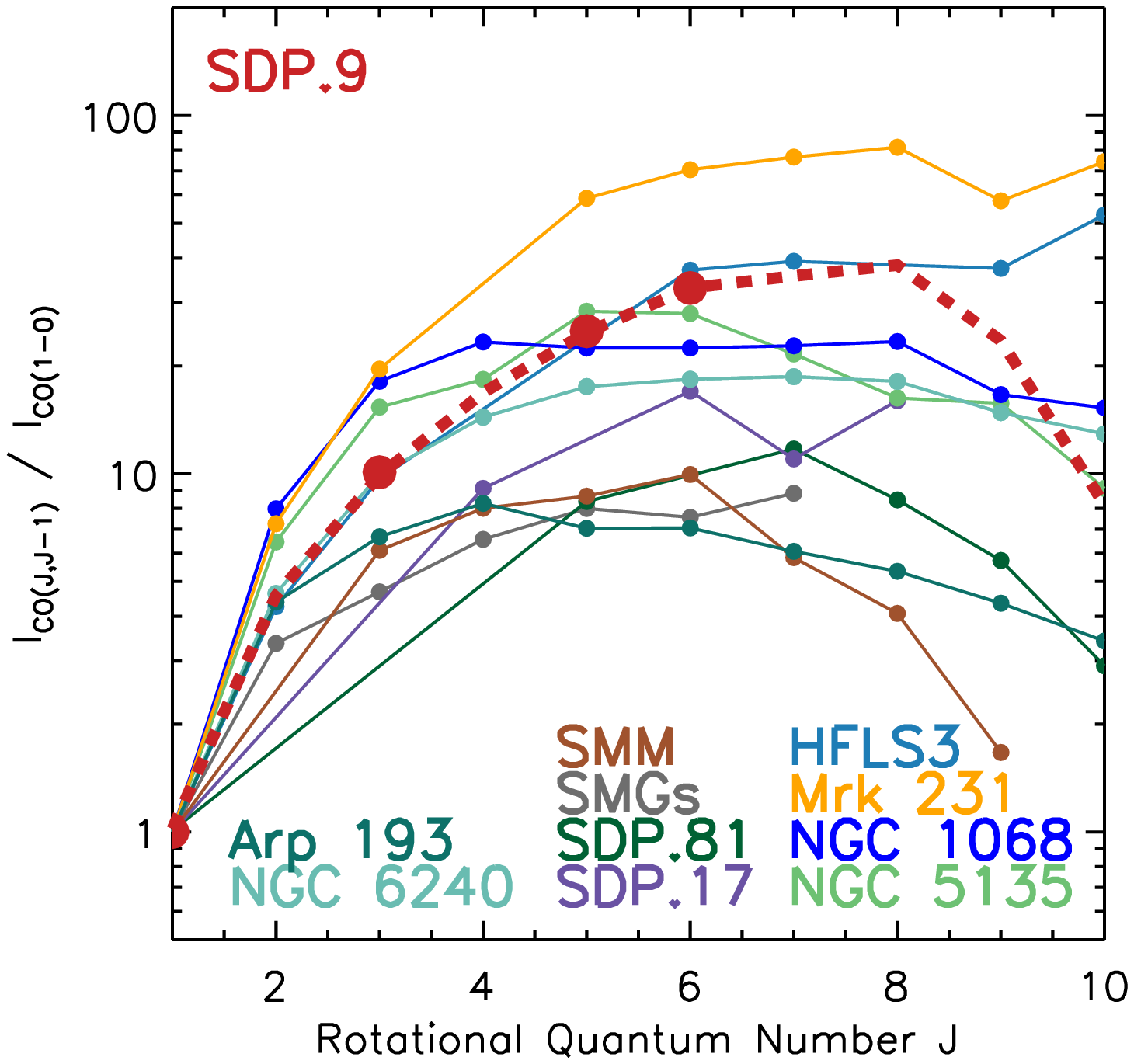} 
\caption{CO spectral line distribution (SLED) of SDP.9 (red), with velocity-integrated line fluxes in units of ${\rm Jy \, km \, s^{-1}}$. Red points are the line fluxes included in Table \ref{table_flux_values}, whereas the red dashed curve is the best-fit model. We also include the CO SLED for other lensed ULIRGs at high redshift \citep{Lupu2012ApJ...757..135L,Vlahakis2015ApJ...808L...4A,Harris2012ApJ...752..152H,Frayer2011ApJ...726L..22F,Omont2013A&A...551A.115O}, the classical population of submm galaxies \citep{Bothwell2013MNRAS.429.3047B}, the lensed SMG SMM J2135$-$0102 at $z \sim 2.3$ \citep[labelled as SMM,][]{Danielson2011MNRAS.410.1687D}, HFLS3 at $z \sim 6.34$ \citep{Riechers2013Natur.496..329R} and several local (U)LIRGs \citep{Meijerink2013ApJ...762L..16M,Papadopoulos2014ApJ...788..153P,Rosenberg2015ApJ...801...72R}. It can be seen that SDP.9 is a highly excited source (but not as excited as local AGNs such as Mrk 231), whose CO SLED resembles the one for HFLS3.
              }
\label{CO_SLED_SDP9}
\end{figure}


\subsection{The CO SLED}\label{CO_SLED_Section}

In order to analyze the excitation conditions of the molecular gas in SDP.9 we have modeled its CO SLED by using all available CO line data, taken from this and previous work (see Table \ref{table_flux_values}). It should be noted that due to lack of low--$J$ ($J_{\rm up} \leq 3$) $^{12}$CO observations for SDP.11, which trace the bulk of its molecular gas, we only study in this work the CO SLED of SDP.9 (Figure \ref{CO_SLED_SDP9}). The inclusion of the $^{12}$CO(1--0) and $^{12}$CO(3--2) transitions in SDP.9 improves the sampling of the CO SLED in this source with respect to the analysis presented in \cite{Lupu2012ApJ...757..135L}. The $^{12}$CO(3--2) emission had been already presented in \cite{Omont2013A&A...551A.115O} at low spatial resolution, and with a line flux similar to the one reported here from our high-spatial resolution data. For a reference, we also include in Figure \ref{CO_SLED_SDP9} the CO SLED of different sources: the classical SMG population \citep{Bothwell2013MNRAS.429.3047B}, high-redshift lensed ULIRGs \citep{Lupu2012ApJ...757..135L,Vlahakis2015ApJ...808L...4A,Harris2012ApJ...752..152H,Frayer2011ApJ...726L..22F}, HFLS3 at $z \sim 6.34$ \citep{Riechers2013Natur.496..329R}, and local (U)LIRGs \citep{Papadopoulos2014ApJ...788..153P,Rosenberg2015ApJ...801...72R}. The shape of the observed CO SLED reveals a highly excited molecular ISM in SDP.9, much more excited than the average SMG (median redshift of $z \sim 2.3$), SDP.81 at $z \sim 3$, slightly more excited than some local merger/starbursts such as NGC\,6240 \citep{Papadopoulos2014ApJ...788..153P}, and less excited than some local AGNs such as Mrk\,231. The similarity between the CO SLED of SDP.9 and HFLS3 is very remarkable. 

In order to model the CO line ladder in SDP.9 we use the non-LTE radiative transfer code Myradex \footnote{https://github.com/fjdu/myRadex}, with an escape probability of $\beta=\frac{1-e^{-\tau}}{\tau}$ in an expanding spherical geometry. We adopt a CO abundance relative to H$_2$ of $10^{-4}$. The CMB temperature is set to $T_{\rm CMB}= 7.004 \, {\rm K}$. We made a grid of CO line ladders with parameter ranging within: number density of molecular hydrogen $n_{\rm H_2} = 10^2$ -- $10^8$ cm$^{-3}$, kinetic temperature $T_{\rm kin} = 10^{0.5} - 10^3 \, {\rm K}$, and velocity gradient $dv/dr = 1 - 10^3 \, {\rm km \, s^{-1} \, pc^{-1}}$. We then convert the velocity gradient to column density, where the latter is used as an input for Myradex, following \cite{Zhang2014A&A...568A.122Z}.  From the model grid, we calculate the likelihood distributions for all parameter combinations following \cite{Ward2003ApJ...587..171W}. We assume that the CO column length is less than $2 \times$ times the CO emitting size. We marginalize the parameters by integrating the likelihood distribution over all other parameters. The best-fit gives the following results: $n(\rm H_2) = 3.2 \times 10^{5} \, {\rm cm^{-3}}$, $dv / dr = 18.2 \, {\rm km \, s^{-1} \, pc^{-1}}$, $T_{\rm kin} = 10^{1.7} \, {\rm K}$, $N_{\rm H_2} = 2.5 \times 10^{22}$, $M_{\rm H_2} = 1.3 \times 10^{11} \, M_\odot$ and $X_{\rm CO} = 1.0 \, {\rm K \, km \, s^{-1} \, cm^{-2}}$. Therefore, the modeling of its CO SLED suggests that the molecular gas in SDP.9 is likely dominated by a warm and dense component, which is much more excited than the average submm population at higher redshift \citep{Bothwell2013MNRAS.429.3047B}.


\subsection{Dense molecular gas tracers in high-redshift starbursts}

\begin{figure}
\centering
\includegraphics[width=0.45\textwidth]{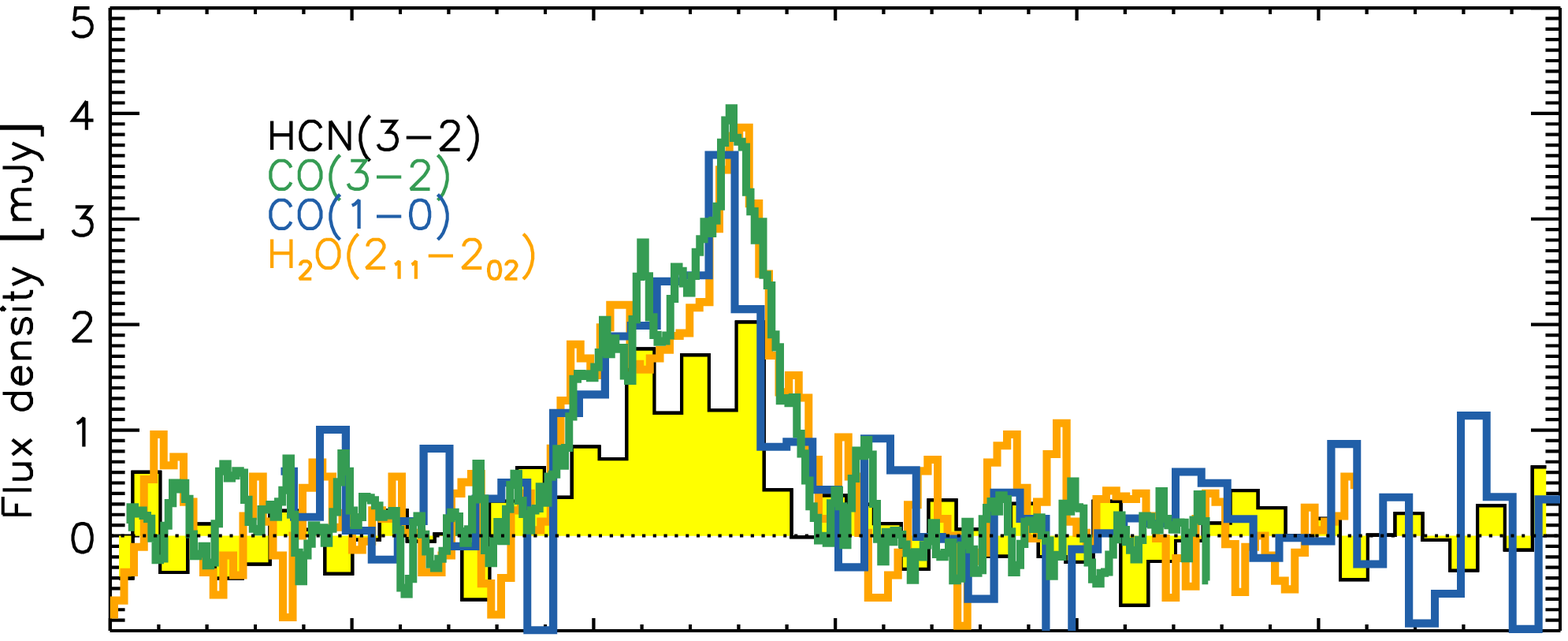} \\
\vspace{-8mm}
\includegraphics[width=0.45\textwidth]{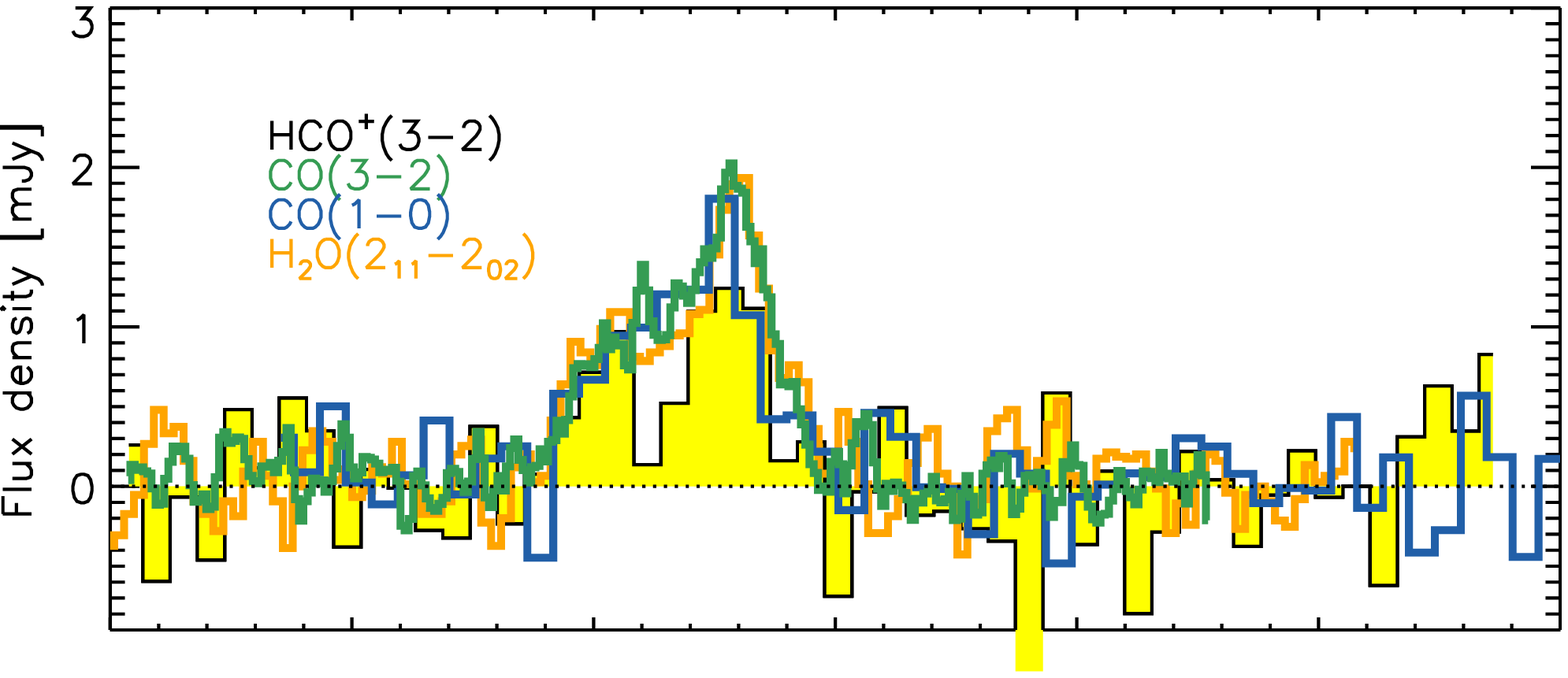} \\
\vspace{-8mm}
\includegraphics[width=0.45\textwidth]{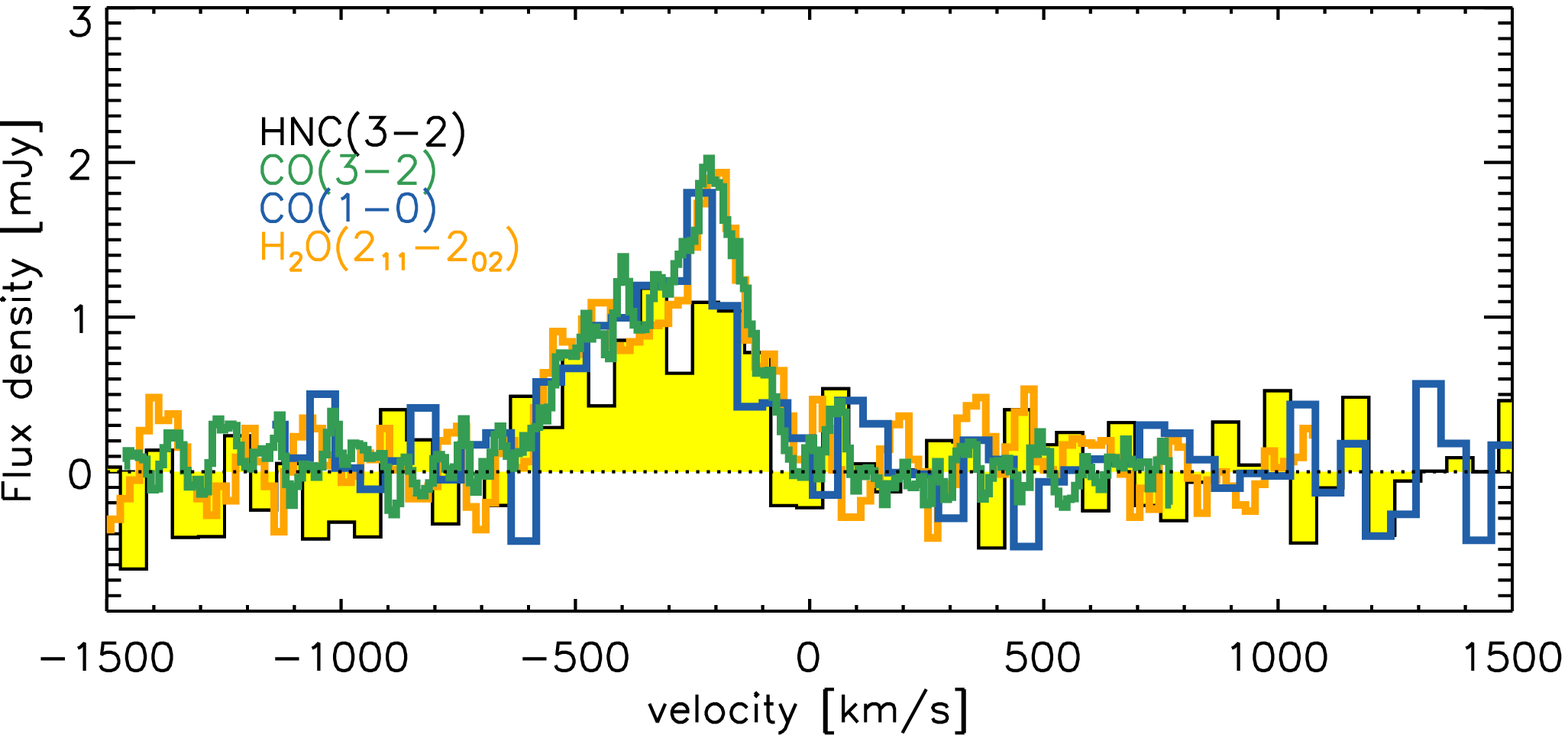}
\caption{ALMA spectra of the detected HCN, HCO$^+$, and HNC lines in SDP.9. We also show the available CO(3--2) (PdBI, this work), CO(1--0) (VLA, this work), and ${\rm H_2O(2_{11}-2_{02})}$ (from \citealt{Omont2013A&A...551A.115O}) transitions in the three panels. For consistency with previous studies of this source, the redshift (see Table \ref{table_flux_values}) has been taken from \citep{Bussmann2013ApJ...779...25B}, although this means that the lines are not center at zero velocity. The CO and water spectra are replicated on each panel to compare their profiles with the dense molecular gas tracer line profiles. All spectra are re-scaled to the HCN(3--2), HCO$^+$(3--2) and HNC(3--2) emission lines so they all fit into the same $y$-axis scale. It should be noted that the HCN(1--0) and HCO$^+$(1--0) spectra are not shown here since the line emission is only detected after collapsing the data cube over the same velocity range where the CO and water lines are detected (see Figure \ref{HCO_mom0_map_JVLA}). It can be seen that the CO and water line profiles are very similar for this source (indicating that the CO and ${\rm H_2O}$ gas are spatially coincident as reported in \cite{Omont2013A&A...551A.115O} and \citealt{Yang2016A&A...595A..80Y}), but slightly different to the dense molecular gas transitions.
              }
\label{image_HST_and_velocitymap}
\end{figure}

\begin{figure}
\centering
\includegraphics[width=0.45\textwidth]{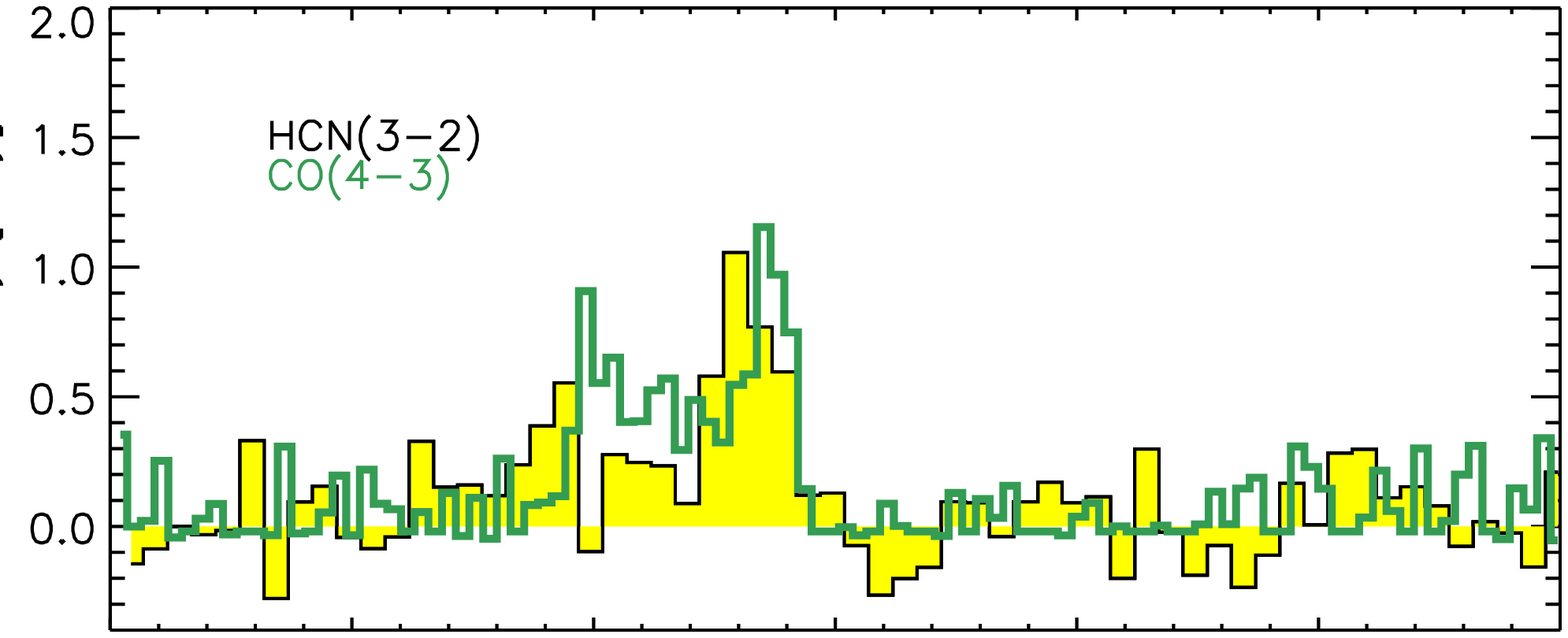} \\
\vspace{-8mm}
\includegraphics[width=0.45\textwidth]{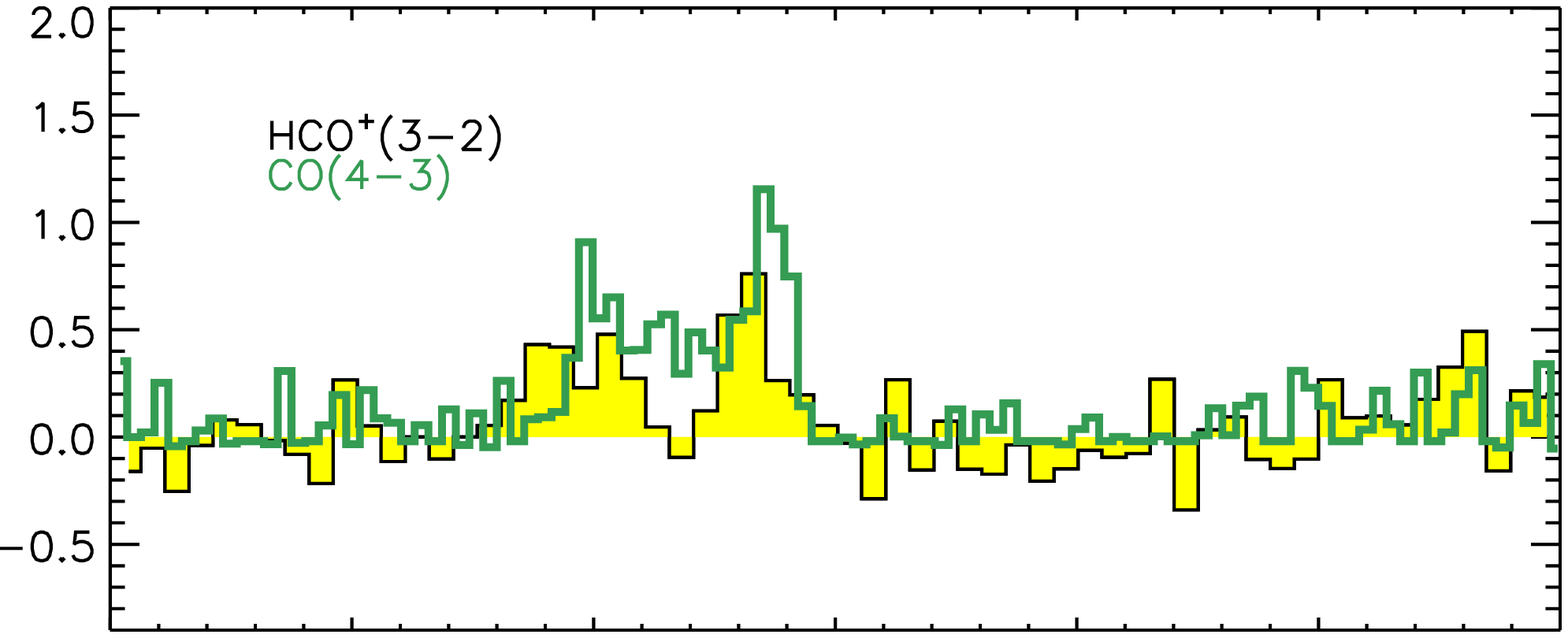} \\
\vspace{-8mm}
\includegraphics[width=0.45\textwidth]{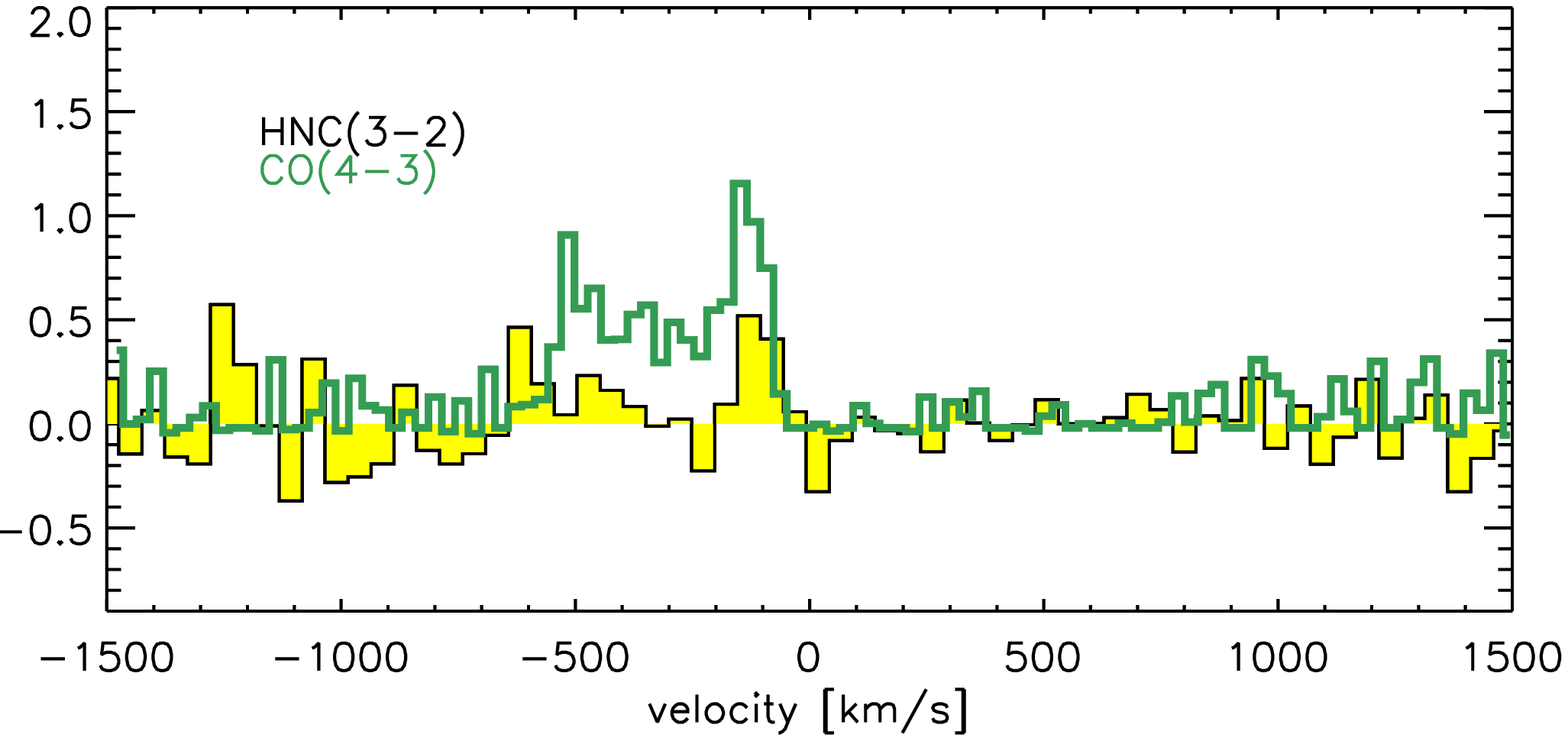}
\caption{Similar to Figure \ref{image_HST_and_velocitymap} but for SDP.11.  
              }
\label{image_HST_and_velocitymap_SDP11}
\end{figure}

\begin{figure}
\centering
\includegraphics[width=0.40\textwidth]{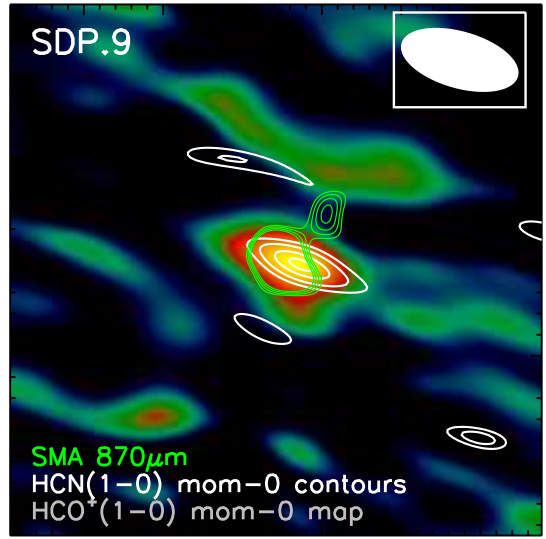}
\caption{Integrated HCN(1--0) and HCO$^+$(1--0) emission in SDP.9 as observed by the VLA. For a reference, we also include the contours of the SMA $870 \, {\rm \mu m}$ continuum emission. All contours are represented from $3 \sigma$ to $4.5 \sigma$, in steps of $0.5 \sigma$. Note that there are no $-3 \sigma$ contours. The synthesized beam of the observations is indicated in the upper-right corner. The map is $7.5''$ on each side, north is up, east is left. Both transitions are detected at $> 4\sigma$ (see line fluxes in Table \ref{table_flux_values}).
              }
\label{HCO_mom0_map_JVLA}
\end{figure}

Figures \ref{image_HST_and_velocitymap} and \ref{image_HST_and_velocitymap_SDP11} show the HCN(3--2), HCO$^+$(3--2) and HNC(3--2) spectra of SDP.9 and SDP.11, respectively, along with previously detected lines in these sources. Thanks to the depth provided by the ALMA observations, all three dense molecular gas transitions are detected in SDP.9, while HCN(3--2) and HCO$^+$(3--2) transitions are detected in SDP.11. The HCN(1--0) and HCO$^+$(1--0) transitions in SDP.9 are detected at $>4\sigma$ after collapsing the data cube over the same velocity range where the CO, ${\rm H_2O}$, HCN(3--2), HCO$^+$(3--2), and HNC(3--2) lines are detected (see Figure \ref{HCO_mom0_map_JVLA}). The velocity-integrated line fluxes (and $3\sigma$ upper limits) of all observed transitions are shown in Table \ref{table_flux_values}. All velocity-integrated line fluxes have been calculated from the zero-moment maps. To calculate the upper limit for the HNC(3--2) transition of SDP.11 we have assumed that its width is the same as the detected CO line and there is no velocity offset with respect to the detected dense gas tracers.

In SDP.9, the width of the HCN(3--2), HCO$^+$(3--2) and HNC(3--2) transitions is similar to the CO and water lines and are also centered at the same velocity. However, it seems (although the relatively low signal to noise prevent robust results) that the line profiles are slightly different. In contrast, despite the molecular gas lines in SDP.11 have similar profiles, they are shifted with respect to each other in about $\sim 100 \, {\rm km \, s^{-1}}$, with the shift being more evident in the HCN(3--2). These results add to the similarities and differences found between dense and CO lines in previous analysis of high-redshift sources. \cite{Riechers2011ApJ...726...50R} found that the HCO$^+$(4--3) transition in the Cloverleaf quasar at $z =2.56$ is narrower than the CO lines, suggesting that the densest gas is more spatially concentrated. \cite{Danielson2011MNRAS.410.1687D} found that the velocity centroid of the HCN(3--2) emission in SMM\,J2135-0102, a lensed SMG at $z \sim 2.3$, is redshifted by approximately $\sim 230 \pm 100\,{\rm km \, s^{-1}}$ with respect to the redshift derived from $^{12}$CO lines.


\subsection{The HCN and HCO$^+$ SLED}

\begin{figure}[!t]
\centering
\includegraphics[width=0.48\textwidth]{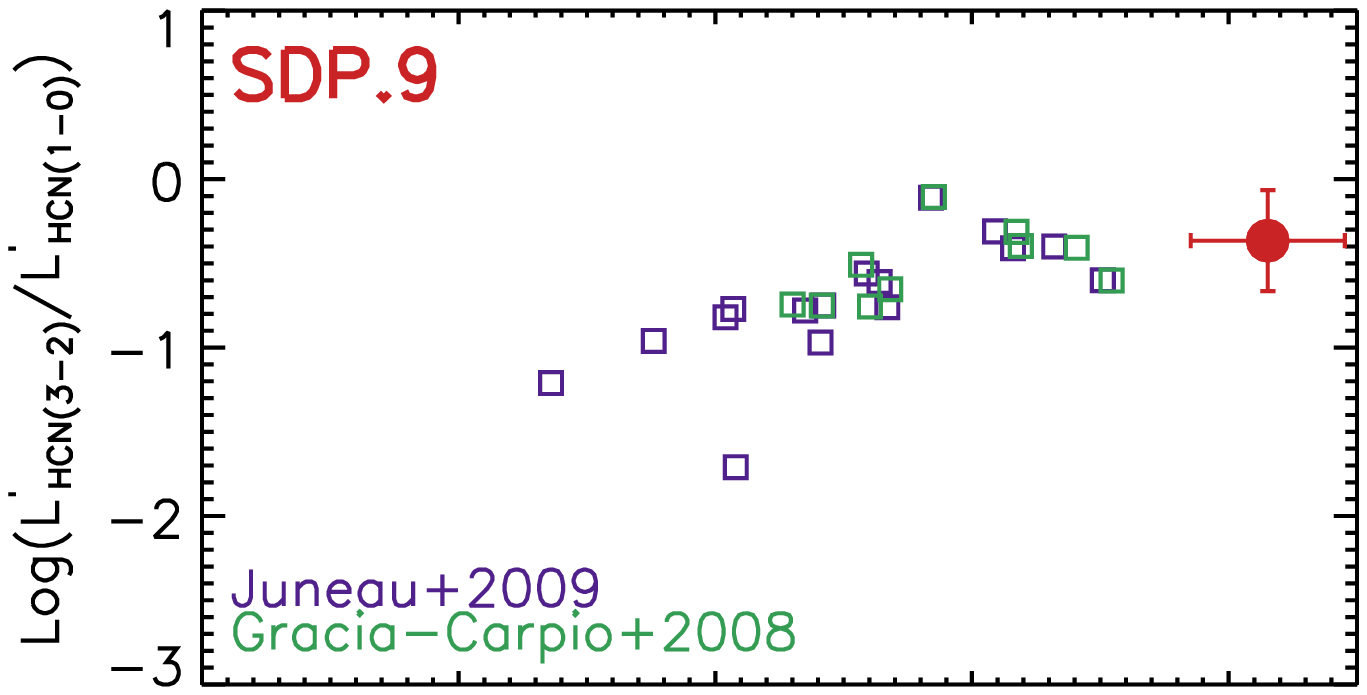} \\
\vspace{-11mm}
\includegraphics[width=0.48\textwidth]{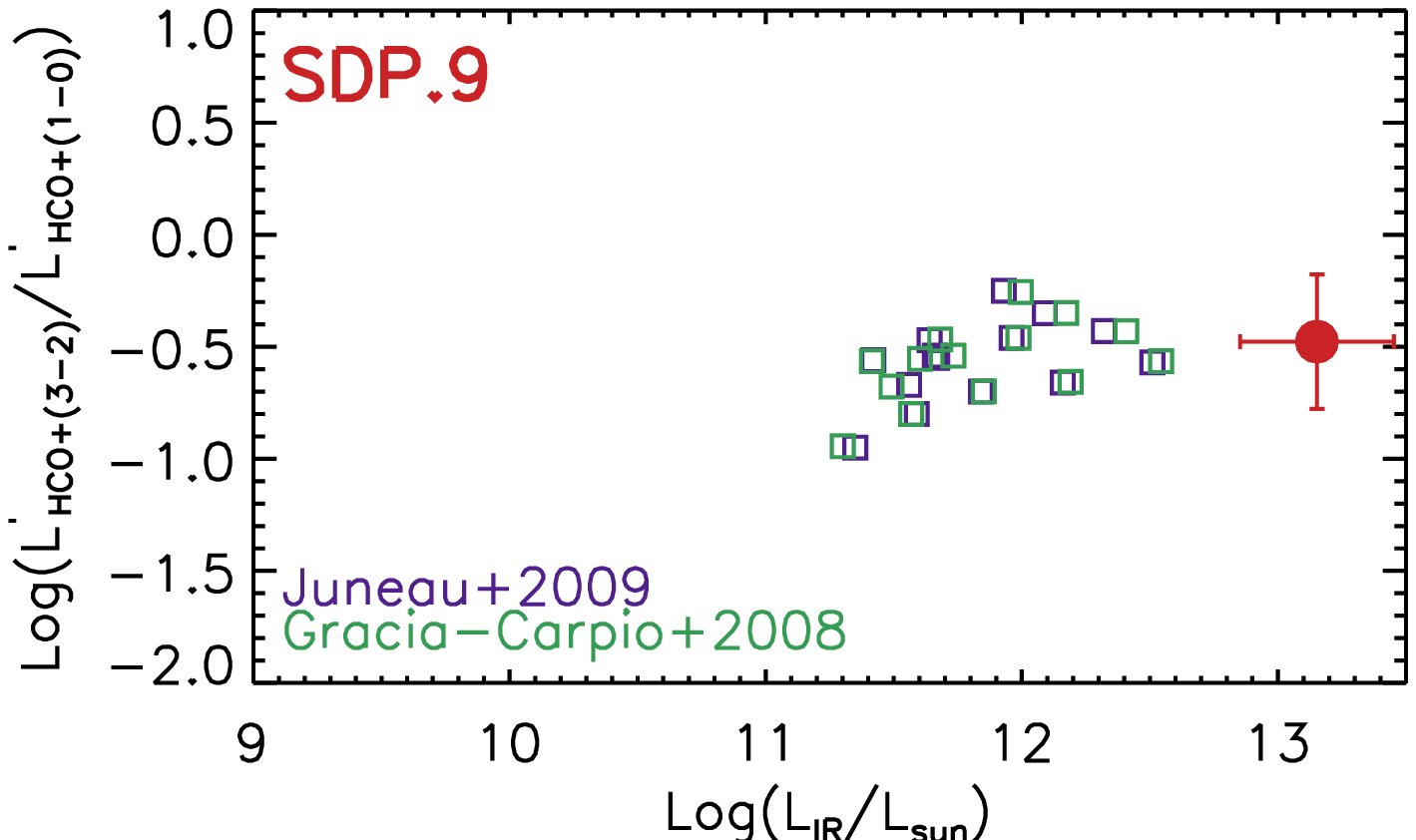}\\
\caption{Ratio between the 3--2 and 1--0 transitions of the dense molecular gas tracers HCN (\emph{upper panel}) and HCO$^+$ (\emph{lower panel}) in SDP.9. These are compared with the values found for local star-forming galaxies \citep{GraciaCarpio2008A&A...479..703G,Juneau2009ApJ...707.1217J}. We see that the line ratios in SDP.9 are compatible with those found for local sources, suggesting similar excitation conditions. Note that SDP.11 is not included in this Figure because its HCN(1--0) and HCO$^+$(1--0) have not been observed.
              }
\label{ratio_HCN_32_HCO_32}
\end{figure}

Figure \ref{ratio_HCN_32_HCO_32} shows the 3--2 to 1--0 luminosity ratio for the HCN and HCO$^+$ transitions in SDP.9. Note that SDP.11 is not included in Figure \ref{ratio_HCN_32_HCO_32} because the 1--0 transitions have not been observed in this source. We see that the HCN(3--2)/HCN(1--0) line ratio in SDP.9 is compatible with the highest values found in local galaxies. This is actually compatible with a trend suggesting higher excitation in more luminous galaxies. On the other hand, the HCO$^+$(3--2)/HCO$^+$(1--0) line ratio is similar to the average value found in local starbursts \citep{GraciaCarpio2008A&A...479..703G,Juneau2009ApJ...707.1217J}, which does not seem to correlate with the total IR luminosity, although no HCO$^+$ transitions have been detected at low luminosity. The line HCN SLED of SDP.9 is also compatible with that found for Arp\,193, a classical starbursts in the local Universe, but suggests a less excited dense molecular gas reservoir than in NGC\,6240 \citep{Papadopoulos2014ApJ...788..153P}. These results suggest that the excitation conditions of the dense molecular gas in some high-redshift sources are not significantly different to what it can be found in the local Universe, despite SDP.9 is more luminous than most sources studied so far.

We have also fitted the CO and HCN SLEDs simultaneously with the single excitation component LVG modeling. To do this we have assumed that both CO and HCN transitions come from the same region with the same excitation conditions. We keep the HCN abundance (relative to ${\rm H_2}$) to be 2$\times 10^{-8}$, which is an average value in local galaxies \citep{Omont2007RPPh...70.1099O,Krips2008ApJ...677..262K}. We derive the likelihood distributions for both grids of HCN and CO, and marginalise the matrix using the same way as in Section \S \ref{CO_SLED_Section}. The best-fit gives the following results: $n(\rm H_2) = 10^{4.5} \, {\rm cm^{-3}}$, $T_{\rm kin} = 10^{1.8} \, {\rm K}$, $dv / dr = 10^{1.48} \, {\rm km \, s^{-1} \, pc^{-1}}$, $N({\rm H_2}) = 10^{22.5}$, $M({\rm H_2}) =10^{11.2} \, M_\odot$ and $X_{\rm CO} = 1.4 \, {\rm K \, km \, s^{-1} \, cm^{-2}}$. We see that the $T_{\rm kin}$ and column density are consistent with the CO-based results. However, the ${\rm H_2}$ volume density is $\rho = 10^{4.5} \, {\rm cm^{-3}}$, lower than the CO-based value. It is likely because the CO excitation is close to a LTE condition that it is not sensitive to density.

\subsection{The $L_{\rm IR}-L'_{\rm dense}$ relations at high redshift}

\begin{figure*}[!ht]
\centering
\includegraphics[width=0.31\textwidth]{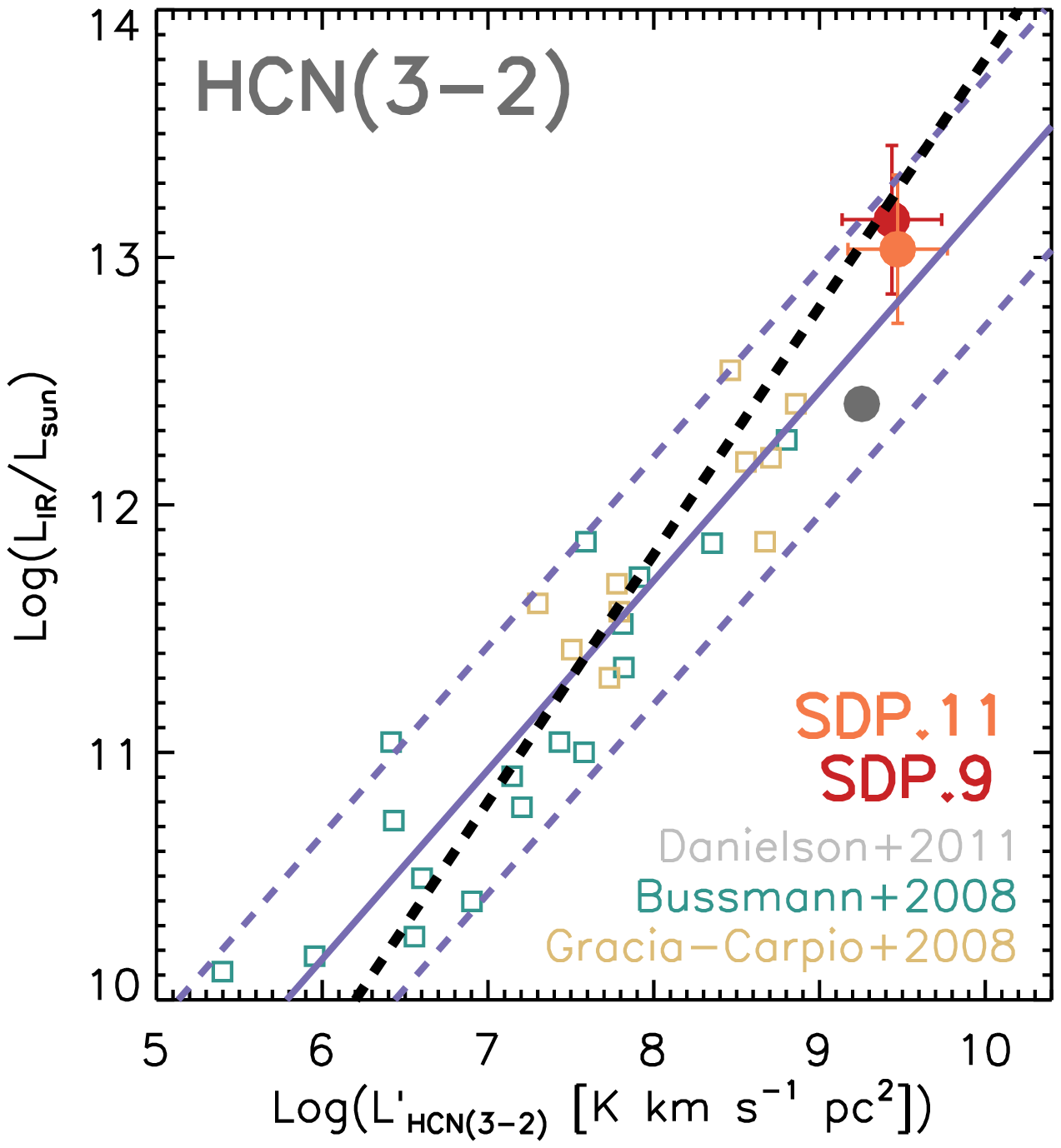} 
\hspace{-17mm}
\includegraphics[width=0.31\textwidth]{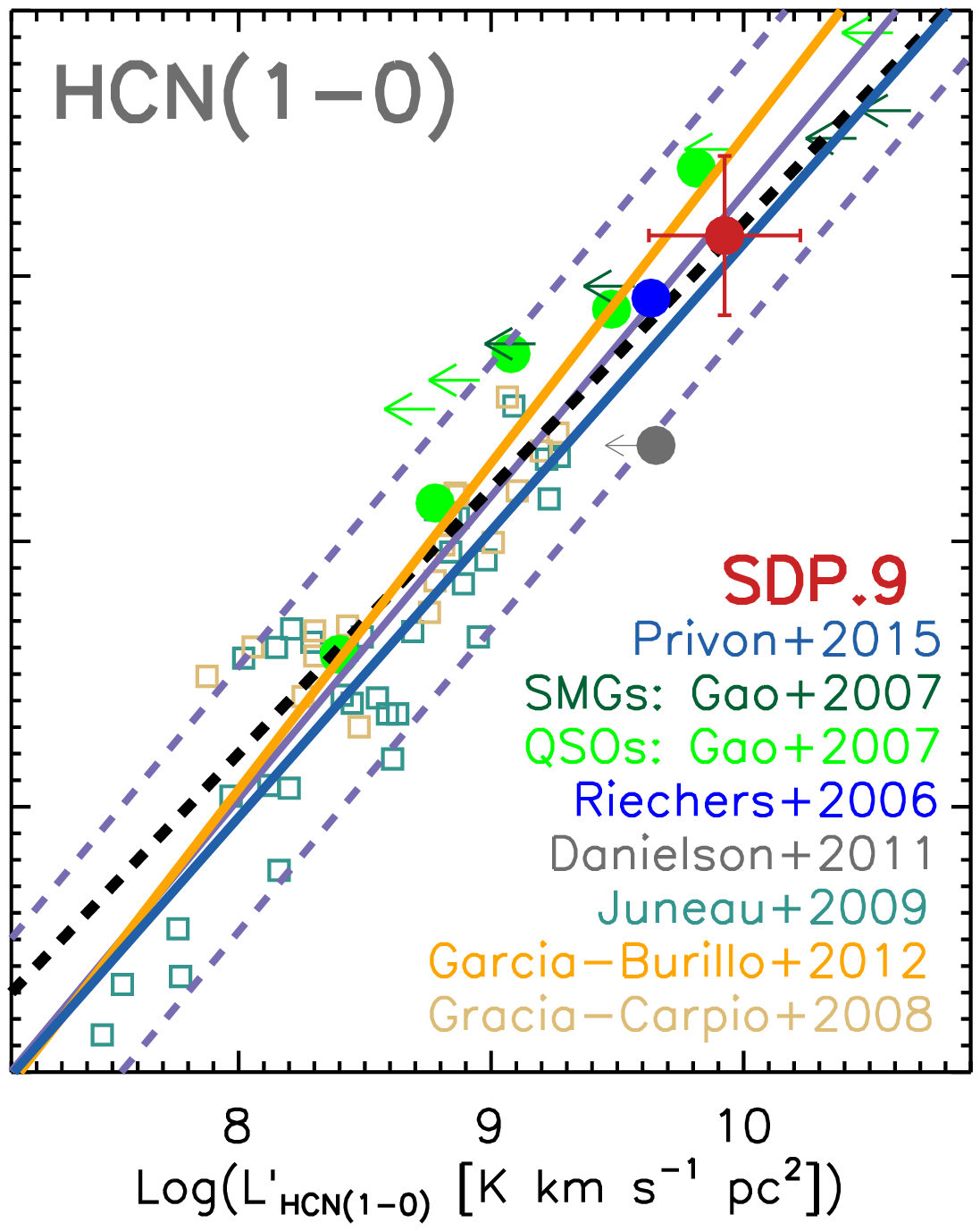} 
\hspace{-17mm}
\includegraphics[width=0.31\textwidth]{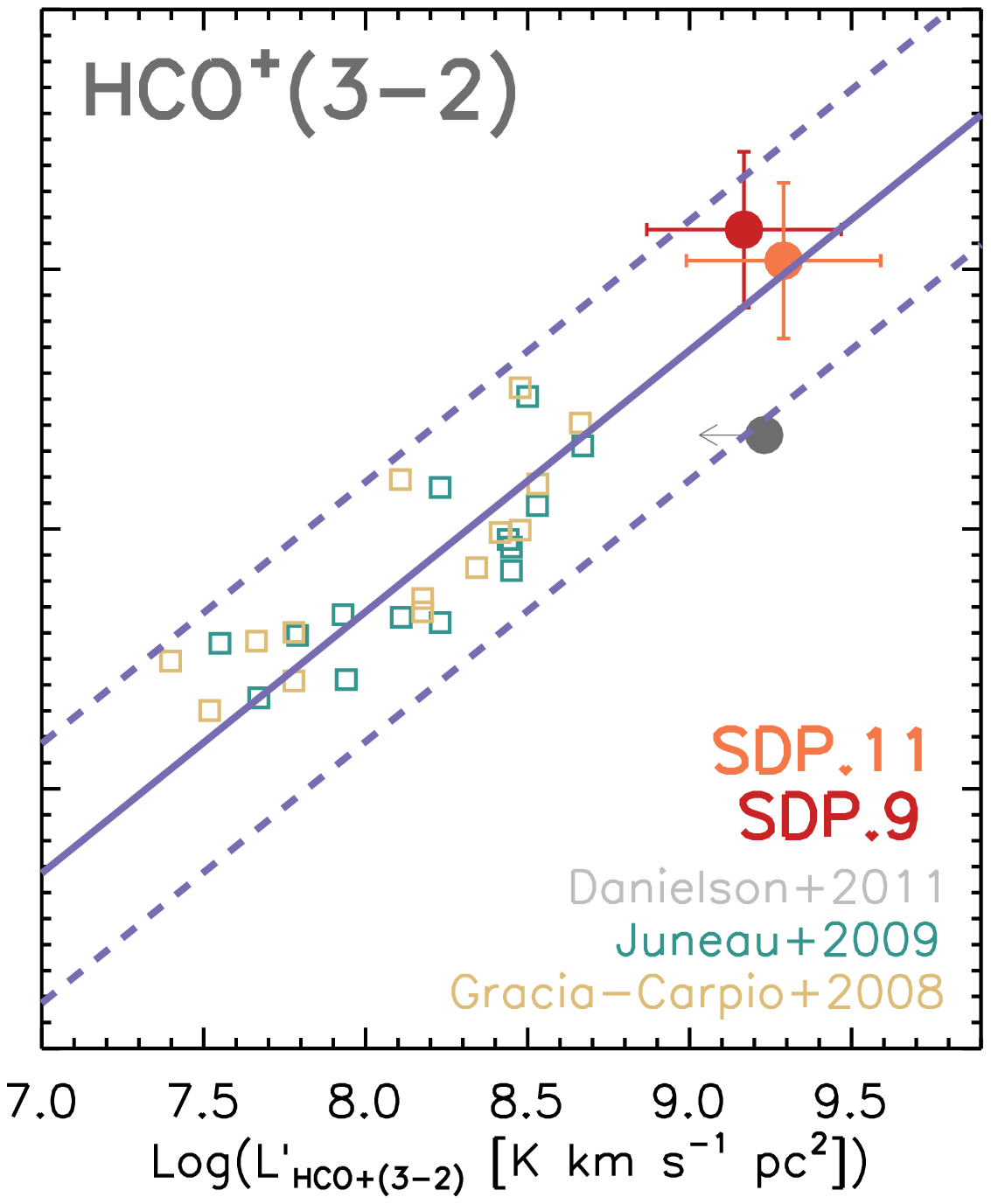} 
\hspace{-17mm}
\includegraphics[width=0.31\textwidth]{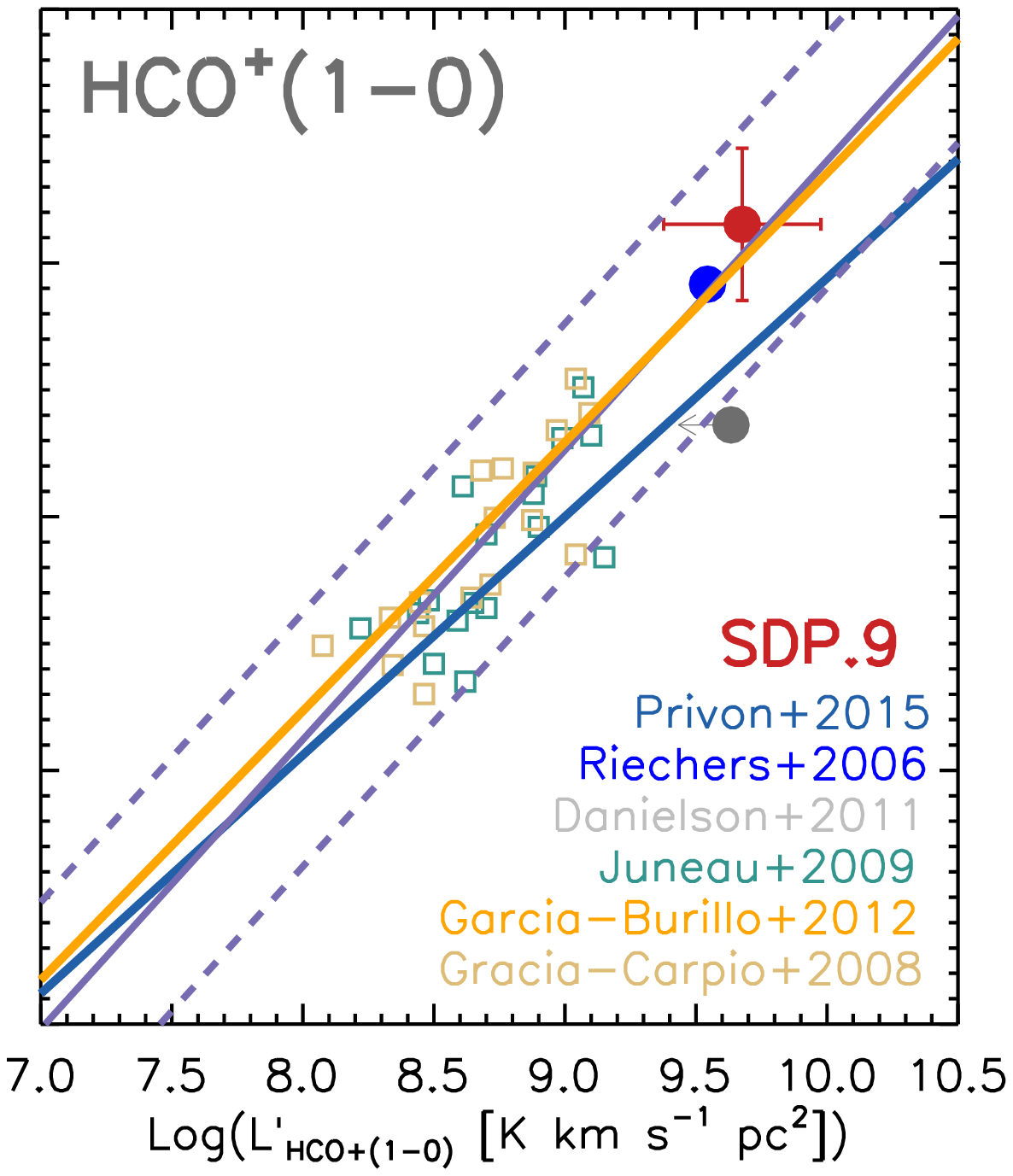} \\
\caption{Relation between the total IR luminosity, $L_{\rm IR}$, and the dense molecular gas luminosities studied in this work for SDP.9 (red dots) and SDP.11 (orange dots). We compare the values found for SDP.9 and SDP.11 with those obtained for different samples of local galaxies \citep{Bussmann2008ApJ...681L..73B,GraciaCarpio2008A&A...479..703G,Juneau2009ApJ...707.1217J} and high-redshift QSOs and SMGs from \cite{Gao2007ApJ...660L..93G}, \cite{Riechers2006ApJ...645L..13R} and \cite{Danielson2011MNRAS.410.1687D}. We caution that \cite{Bussmann2008ApJ...681L..73B} and \cite{Juneau2009ApJ...707.1217J} might have not corrected their measurements for the effect that the beam of their observations did not cover the whole extension of their sources. Purple solid lines represent linear fit to all point on each panel and purple dashed lines represent $\pm 0.5$ dex deviation from the linear fit. We also represent the correlation between $L_{\rm IR}$ and $L_{\rm HCN}$ and $L_{\rm HCO^+}$ derived in \cite{GarciaBurillo2012A&A...539A...8G} with solid orange lines. For a reference, we include with a black dashed line a relation between $L_{\rm IR}$ and $L'_{\rm HCN(1-0)}$ with a slope equal to one, expected for ULIRGs, see discussion in \cite{Riechers2007ApJ...671L..13R}. Although with a relatively large scatter, all points for local and high-redshift starbursts can be fitted with a single linear relation. Therefore, the strong SFR in our galaxies is accompanied by a more massive dense molecular gas reservoir in the same proportion as in the local Universe. It should be noted that because we are assuming the same magnification factor for the total IR and dense gas luminosities, the location of our two sources in this diagram with respect to local trends is only affected by differential amplification, but not by the absolute value of the amplification factor.              }
\label{LIR_LHCN_relations}
\end{figure*}

Figure \ref{LIR_LHCN_relations} compares the total IR and dense molecular gas luminosities of SDP.9 and SDP.11 with a sample of local and high-redshift starbursts and QSOs taken from the literature \citep{Riechers2006ApJ...645L..13R,Gao2007ApJ...660L..93G,Greve2006AJ....132.1938G,Bussmann2008ApJ...681L..73B,GraciaCarpio2008A&A...479..703G,Juneau2009ApJ...707.1217J,Danielson2011MNRAS.410.1687D,GarciaBurillo2012A&A...539A...8G}. It should be pointed out that the line fluxes reported in \cite{Bussmann2008ApJ...681L..73B} and \cite{Juneau2009ApJ...707.1217J} do not seem to take into account the fact that their galaxies are larger than the beam size of their single-dish observations, thus missing a fraction of the dense gas line fluxes. We see that the correlation between IR and line luminosity for both low-redshift galaxies and SDP.9 and SDP.11 can be described by a single relation. Therefore, we find no clear evidence of enhanced SFR at a given line luminosity in SDP.9 and SDP.11 when compared to local LIRGs/ULIRGs. This is mainly because the correlation between $L_{\rm IR}$ and $L_{\rm dense}$ for local galaxies is not specially tight, but instead has a dispersion of $\pm 0.5 \, {\rm dex}$ (see Figure \ref{LIR_LHCN_relations}). 

The detected HCN(3--2) emission in SMM\,J2135-0102 reported by \cite{Danielson2011MNRAS.410.1687D} has line and IR luminosities which are also in agreement with the spread of the local correlation, also supporting the fact that a single relation can be applied at low and high redshift. \cite{Gao2007ApJ...660L..93G} presented HCN(1--0) observations of two SMGs and two QSOs and compiled previous HCN(1--0) detections and upper limits from the literature to conclude that high-redshift sources systematically lie above FIR/HCN correlation for nearby galaxies by about a factor of 2. This does not seem to be in agreement with our results for SDP.9 and SDP.11. However, it should be pointed out that the sample used in \cite{Gao2007ApJ...660L..93G} has a significant contribution of high-redshift QSOs (even at redshift $z \sim 6$) whose dense molecular gas properties does not necessarily have to be similar to the classical population of high-redshift SMGs (median redshift of $z \sim 2.3$ -- \citealt{Chapman2005}) or $z \sim 1$ ULIRGs as those studied here. Furthermore, the total IR luminosities in the former sources might be affected by the AGN, so not all the total IR luminosity is due to star formation. If that was actually the case, and only the total IR luminosity due to star formation was considered, the point would be closer to the local relation. On the other hand, \cite{Riechers2006ApJ...645L..13R} reported detections of HCN(1--0) and HCO$^+$(1--0) in the Cloverleaf quasar at $z = 2.56$, and their points are also compatible with the local relation (see Figure \ref{HCN_HCO_HNC_lineratios_fig}). 

The fact that, as suggested by our results, the same relation between total IR and dense gas luminosities can be applied to galaxies both at low and high redshift might indicate that some high-redshift sources are more luminous (have higher SFR) just because they have more dense molecular gas. 


It has been obtained in different previous work that the relation between $L_{\rm IR}$ and $L'_{\rm HCN}$ from $J \geq 1$ has a slope equal to one, meaning that all those dense gas tracers are correlated with the SFR and indicating that the SFR in the dense gas is not likely affected by the free-fall time \citep{Zhang2014ApJ...784L..31Z}. Actually, because the free-fall time is related to the density by $\tau_{\rm ff} \propto n^{-0.5}$, it would be lower for higher-$J$ transitions, but still a similar $L_{\rm IR} - L_{\rm dense}$ relation is seen for $^{12}$CO(1--0) and $^{12}$CO(4--3), HCN(1--0), HCN(4--3) and other dense gas tracers \citep{Wu2010ApJS..188..313W,Reiter2011ApJS..195....1R,Zhang2014ApJ...784L..31Z}. We have represented the a liner correlation with slope equal to unity with the black dashed lines in Figure \ref{LIR_LHCN_relations}. It can be seen that, for the HCN(1-0) transition, and within the uncertainties, the point for SDP.9 is still compatible with a linear relation, and actually most previous fits to the $L_{\rm IR} - L_{\rm HCN(1-0)}$ are compatible with a linear relation \citep{Privon2015ApJ...814...39P,GarciaBurillo2012A&A...539A...8G}. The same seems to be true for the HCN(3-2) transition for both SDP.9 and SDP.11, although the spread of the points for local galaxies prevents a robust determination of the slope. 


\subsection{The dense gas depletion time}

The dense molecular line luminosities can be combined with the total IR luminosity of SDP.9 to determine its star formation efficiency and dense gas depletion time. We note that we do this calculation only for SDP.9 because we have not observed the dense $J = 1-0$ transitions in SDP.11. The dense to total IR luminosity ratio (a proxy of the star-formation efficiency of the dense gas phase) in SDP.9 is $L'_{\rm HCN(1-0)} / L_{\rm IR} \sim 0.7\times 10^{-3} \, L'/L_\odot$ (implying a dense gas depletion time of $\tau_{\rm dep} \sim 40 \, {\rm Myr}$, assuming $M_{\rm dense} [M_\odot] = 10 \, L^{'}_{\rm HCN(1-0)} \, {\rm [K \, km \, s^{-1} \, pc^2]}$, see \citealt{GarciaBurillo2012A&A...539A...8G}) is compatible (within the scatter) with the average value reported for LIRGs in the local Universe \citep{Privon2015ApJ...814...39P,GarciaBurillo2012A&A...539A...8G}, where no significant correlation has been found between the dense gas depletion time and the total IR luminosity \citep{Gao2004ApJ...606..271G}. Actually, a large scatter of $L'_{\rm HCN(1-0)} / L_{\rm IR}$ values have been found, interpreted by \cite{Privon2015ApJ...814...39P} as a lack of a simple link between $L'_{\rm HCN(1-0)}$ and the mass of the dense gas directly associated by star formation.


\subsection{HCN, HCO$^+$ and HNC line ratios at high redshift}

\begin{figure}[!t]
\centering
\includegraphics[width=0.48\textwidth]{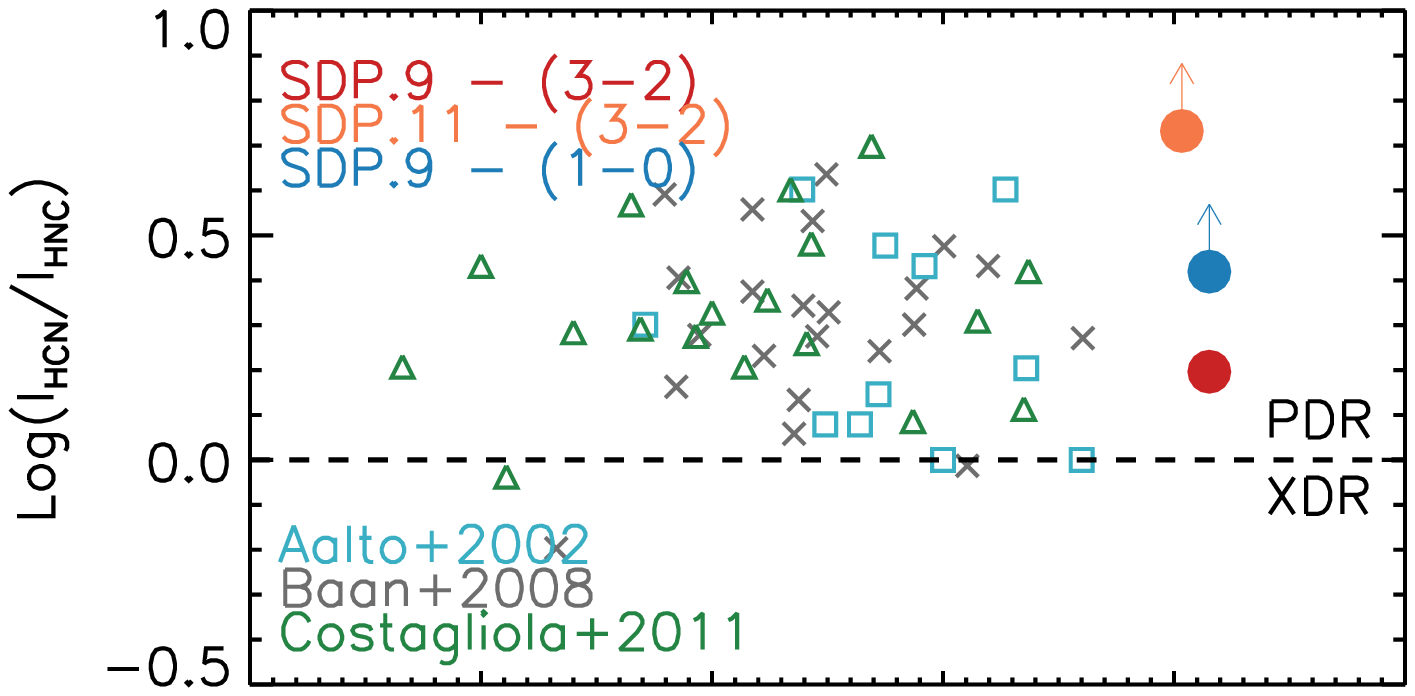} \\
\vspace{-11mm}
\includegraphics[width=0.48\textwidth]{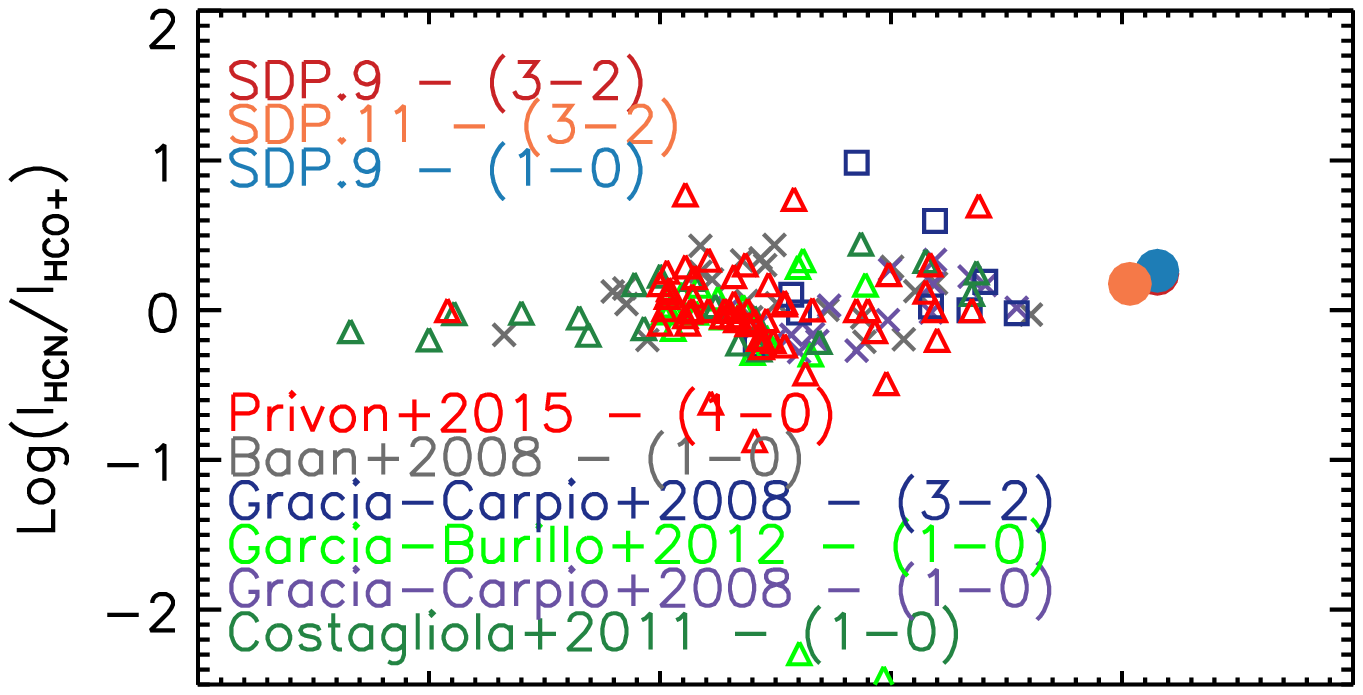}\\
\vspace{-11mm}
\includegraphics[width=0.48\textwidth]{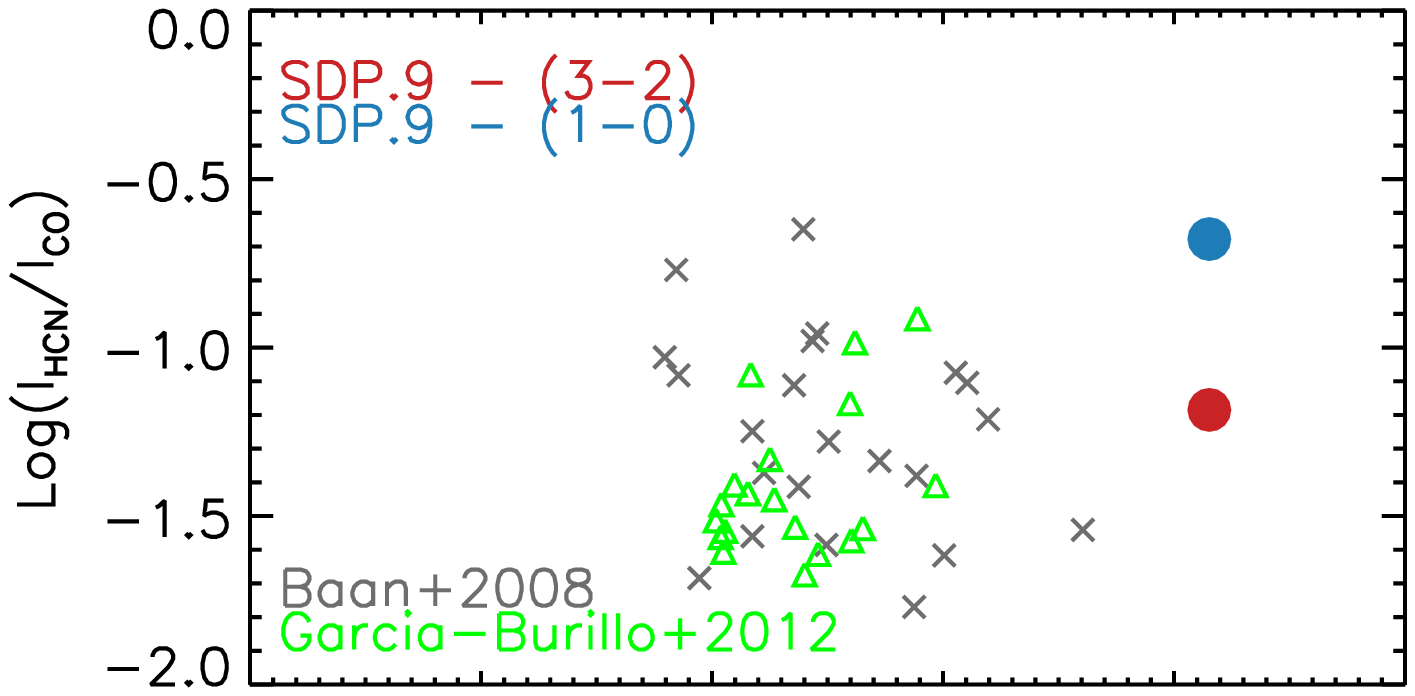} \\
\vspace{-11mm}
\includegraphics[width=0.48\textwidth]{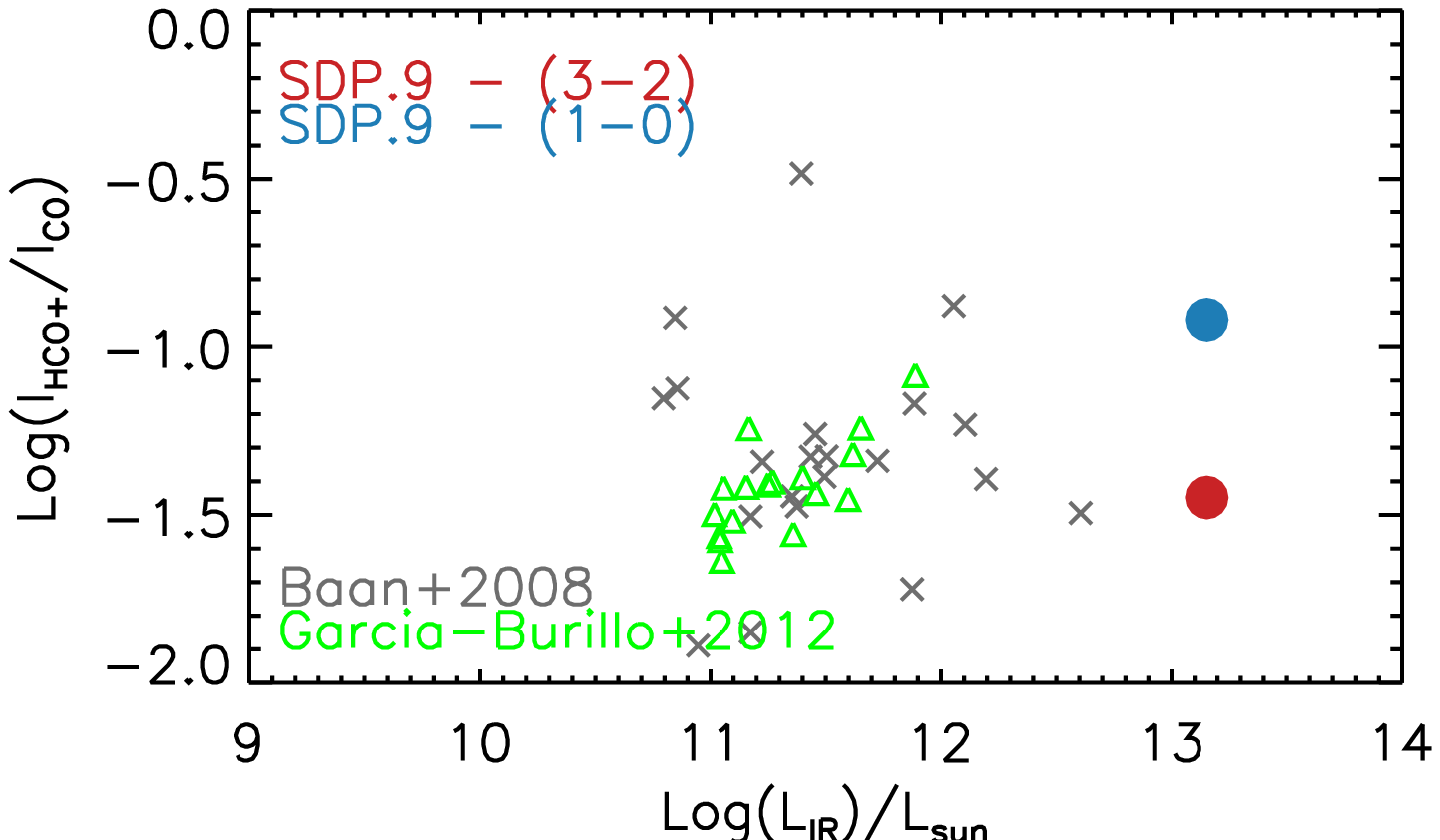} \\
\caption{Line flux ratios for SDP.9 and SDP.11 compared with trends found in local sources. Each panel corresponds to a different line ratio, as indicated by each legend. Note that the similarity in the HCN/HCO$^+$ for SDP.9 makes the red dot appear behind the blue one.
              }
\label{HCN_HCO_HNC_lineratios_fig}
\end{figure}


\subsubsection{The HCN/HNC ratio}

The upper panel of Figure \ref{HCN_HCO_HNC_lineratios_fig} compares the HCN(1--0)/HNC(1--0) and HCN(3--2)/HNC(3--2) line ratios in SDP.9 and SDP.11 with those for a sample of local sources from \cite{Aalto2002A&A...381..783A,Baan2008A&A...477..747B} and \cite{Costagliola2011A&A...528A..30C}. It has been suggested that sources can be divided into PDR or XDR depending on the value of the HCN(1--0)/HNC(1--0) ratio \citep{Meijerink2007A&A...461..793M,Baan2008A&A...477..747B,Loenen2008A&A...488L...5L}. XDRs (along with pumping of the rotational levels through the mid-infrared continuum) is actually one of the plausible explanations for over luminous HNC(1-0) emission in galaxies with warm molecular gas \citep{Aalto2007A&A...464..193A}. According to the models of \citep{Meijerink2007A&A...461..793M}, the high HCN(1--0)/HNC(1--0) ratios found for SDP.9 and SDP.11 could be only explained by PDR dominated regions \citep[see also][]{Loenen2008A&A...488L...5L}. This is similar to the dense gas properties of local ULIRGs \citep{Costagliola2011A&A...528A..30C}, which show an average HCN(1--0)/HNC(1--0) ratio of $\sim 2$. Actually, as shown in Figure \ref{HCN_HCO_HNC_lineratios_fig}, the HCN(1--0)/HNC(1--0) ratio of most local (U)LIRGs are similar to that found for SDP.9. On the other hand, the HCN(3--2)/HNC(3--2) ratios (or upper limit) found for SDP.9 and SDP.11 could be explained by either XDR or PDR models, since the HCN/HNC ratio for mid-$J$ transitions can be also enhanced in XDRs \citep{Meijerink2007A&A...461..793M}.

As shown in \cite{Loenen2008A&A...488L...5L}, the high HCN(1--0)/HNC(1--0) ratios found for local galaxies, and also for SDP.9, cannot be explained only with models in which PDR or XDR dominates \citep[see also][]{Costagliola2011A&A...528A..30C}. Instead, they show that the effect of mechanical heating (which can be produced by dissipating supernova shocks) needs to be considered \citep[see also][]{Kazandjian2012A&A...542A..65K,Izumi2016ApJ...818...42I}. \cite{Loenen2008A&A...488L...5L} proposed that the PDR sources can be divided into two groups: one small group that can be modeled with classical PDR chemistry and whose HCN/HNC ratios are around unity (which is not the case of SDP.9) and a second group (larger than the other one) with lower densities that are heated by mechanical heating. SDP.9 would fall into the second group due to its high HCN(1--0)/HNC(1--0) ratio and, therefore, mechanical heating should be also included as a possible source of enhanced HCN emission with respect to HNC.


\subsubsection{The HCN/HCO$^+$ ratio}\label{HCN_HCO_ratio_section}

The second panel of Figure \ref{HCN_HCO_HNC_lineratios_fig} shows the HCN(1--0)/HCO$^+$(1--0) and HCN(3--2)/HCO$^+$(3--2) line ratios for SDP.9 and SDP.11 in comparison with the values for a sample of local (U)LIRGs from \cite{GraciaCarpio2008A&A...479..703G,Baan2008A&A...477..747B,Costagliola2011A&A...528A..30C} and \cite{GarciaBurillo2012A&A...539A...8G}. It can be seen that, for both $J = 1-0$ and $J = 3-2$ transitions, the HCN emission is brighter than HCO$^+$ in SDP.9 and SDP.11, and that the line ratios are compatible to the values found for luminous starbursts in the local Universe (although not as luminous as SDP.9 or SDP.11). This might indicate that the dense molecular gas can have similar properties at high and low redshift, at least for a subsample of high-redshift galaxies.

It has been proposed in the literature that the HCN/HCO$^+$ ratio can be used as a tool to distinguish between AGNs and starbursts dominated by star formation. \cite{Kohno2001ASPC..249..672K} suggested that the enhanced HCN in some Seyfert galaxies (with HCN/HCO$^+ > 2$ -- so higher than the values found for SDP.9 and SDP.11) could originate from X-ray-irradiated dense obscuring tori, as it happens in NGC\,1068, NGC\,1097, and NGC\,5194. The higher HCN/HCO$^+$ ratios in AGNs have been attributed to an enhancement of the HCN line in the XDR surrounding the AGN \citep{Lepp1996A&A...306L..21L}. However, \cite{Costagliola2011A&A...528A..30C} claimed that the HCN/HCO$^+$ ratio is not a reliable tracer of XDRs. Furthermore, \cite{Privon2015ApJ...814...39P} recently showed that the HCN/HCO$^+$ ratio in galaxies dominated by star formation can be as high as those found in AGNs. Actually, \cite{Privon2015ApJ...814...39P} showed that neither the hardness of the X-ray spectrum nor the total X-ray luminosity correlate with the HCN/HCO$^+$ ratio, suggesting that XDRs are not always the major driver in enhancing the HCN(1--0) emission in luminous starbursts, possibly because the XDRs are spatially disconnected from the regions that dominate the global line luminosity. 

The HCN emission in molecular clouds can also be enhanced by mid-IR pumping of a $14 \, \mu {\rm m}$ vibrational band \citep{Aalto1995A&A...300..369A}. Additionally, it has been also claimed that the HCN/HCO$^+$ ratio can be also enhanced in compact environments, since in this case self-absorption is more likely to happen in HCO+ than in HCN \citep{Aalto2015A&A...584A..42A}. On the other hand, the HCO$^+$ emission can be enhanced (so decreasing the HCN/HCO$^+$ ratio) by mid-IR pumping via the $12 \, {\rm \mu m}$ vibrational band \citep{GraciaCarpio2006ApJ...640L.135G} and its abundance is affected by cosmic rays and the ratio of cosmic ionization rate and gas density \citep{Riechers2006ApJ...645L..13R}. Both HCN and HCO$^+$ are sensitive to the cosmic-rays-produced abundance of ${\rm H_3}^+$, but HCO$^+$ is an ion and, therefore, also very sensitive to the ambient free electron abundance. Even a small increase of the free electron fraction can lead to a severe HCO$^+$ depletion. Therefore, strong cosmic-ray ionising radiation increases the number of $H_3^+$ to form HCO$^+$, but also increases the presence of free electrons that destroy HCO$^+$ while HCN remains \citep[see discussion in][]{Papadopoulos2007ApJ...656..792P}. 

From all previous work we conclude that the number of processes affecting the observed HCN/HCO$^+$ line ratio is quite significant. Furthermore it seems that there is no consensus about the reliability of using the HCN/HCO$^+$ line ratio as a tracer of the presence of XDRs and, therefore, we do not attempt to draw any conclusion on the nature of SDP.9 and SDP.11 from this line ratio. However, it is very remarkable that the HCN/HCO$^+$ ratios for both SDP.9 and SDP.11 are similar to the values found in local (U)LIRGs. This might indicate that, whatever the processes governing that line ratio, they can be present and affecting that line ratio both at low and high redshift.

\subsection{Comparing high- and low-density molecular gas tracers}

The ratio between the HCN and CO low-$J$ transitions is arguably the cleanest indicator of the the fraction of the total gas reservoir residing in the dense phase, and it has been proposed to be a unique tool to explore the star-formation mode of star-forming galaxies: isolated disk versus merger-driven starbursts \citep{Papadopoulos2012ApJ...757..157P}. Figure \ref{HCN_HCO_HNC_lineratios_fig} includes a comparison between the high-density (HCN and HCO$^+$ emission) and low-density (CO emission) molecular gas tracers for SDP.9, the only of the two sources studied in this work with detections in all those transitions both at $J = 3-2$ and $J = 1-0$. While the ratio for the $J = 3-2$ transitions (both HCN/CO and HCO$^+$/CO) and the $J = 1-0$ HCO$^+$/CO are in very good agreement with the values found for local galaxies \citep{Baan2008A&A...477..747B}, the $J = 1-0$ HCN/CO ratio in SDP.9 is higher than most of the local sources. This indicates that SDP.9 contains a larger dense molecular gas reservoir with respect to the total molecular gas than those found in local galaxies. Since SDP.9 is more IR luminous than most galaxies studied so far in the local Universe, their higher dense gas fraction might be associated to its higher SFR. 

Assuming the relation between the dense molecular gas mass and the HCN(1--0) luminosity $M_{\rm dense} \sim 10 \times L'_{\rm HCN(1-0)} \, {\rm [K\, km\, s^{-1} \, pc^{-2}]}$ \citep{Gao2004ApJS..152...63G,GarciaBurillo2012A&A...539A...8G} we obtain $M_{\rm dense} \sim 8.4 \times 10^{10} \, M_\odot$, which is slightly smaller than the total molecular gas mass estimated from the LVG modelling of the CO SLED ($M_{\rm H_2} \sim 1.3 \times 10^{11} \, M_\odot$, see \S \ref{CO_SLED_Section}). This suggests that a very large fraction of the molecular gas in SDP.9 is in the form of dense gas. The $L'_{\rm HCN(1-0)} / L'_{\rm CO(1-0)} \sim 0.37$ found for SDP.9 is similar to the values found for the most luminous sources in the local Universe \citep{Gao2004ApJ...606..271G,GarciaBurillo2012A&A...539A...8G} and it is also compatible with a correlation indicating that more luminous galaxies have higher dense gas fractions (or equivalently higher $L'_{\rm HCN(1-0)} / L'_{\rm CO(1-0)}$ ratios).

We should point out that there are several uncertainties in the derivation of the dense molecular gas mass. As discussed in \cite{Papadopoulos2014ApJ...788..153P}, the $\alpha_{\rm HCN}$ factor might be lower than typically assumed in extreme starbursts if their dense gas is much warmer and/or in unbound states. Additionally, the HCN/CO line ratio traces the dense gas mass fractions only if both lines are collisionally excited. In strong starbursts, HCN can also be excited by mid-IR IR pumping (see also \S \ref{HCN_HCO_ratio_section}) thus producing enhanced HCN/CO line ratios without large amounts of dense molecular gas. Furthermore, the possible presence of an AGN in SDP.9 might also produce enhanced HCN luminosity \citep{Privon2015ApJ...814...39P}.

As mentioned above, observations of local galaxies conclude that the $J = 1-0$ HCN/CO line ratio is an excellent way to discriminate between different star-formation modes. Compact, merger-driven, extreme starbursts are associated to high $J = 1-0$ HCN/CO ratios, whereas lower values are found for isolated star-forming disks with more extended star formation \citep{Solomon1992ApJ...387L..55S,Gao2004ApJS..152...63G}. SDP.9 might belong to the former group as expected from its high IR luminosity \citep{Engel2010ApJ...724..233E} and, in fact, the $L'_{\rm HCN}/L'_{\rm CO}$ luminosity ratio of SDP.9 is compatible with it being a merger-driven starburst.

According to the models presented in \citep{Meijerink2007A&A...461..793M} the HCN/CO ratio in SDP.9 suggests a density of about $\sim 2-3 \times 10^5 \, {\rm cm^{-3}}$, which is compatible with the density derived from the LVG modeling of the CO SLED, $n_{\rm H_2} \sim 3.2 \times 10^{5}\, {\rm cm^{-3}}$ (see \S \ref{CO_SLED_Section}).

%

\section{Conclusions}\label{conclu}

In this work we have presented ALMA and VLA observations of dense molecular gas tracers in two lensed ULIRGs at $z \sim 1.6$ selected from the {\it H}-ATLAS survey. The main conclusions of our work are:

\begin{enumerate}

\item We have detected all HCN(3--2), HCO$^+$(3--2), HCN(1--0), and HCO$^+$(1--0) transitions in our two sources. Additionally, we have detected HNC(3--2) emission in SDP.9. This clearly highlights the power of ALMA and VLA to carry out studies of the dense molecular gas in high-redshift dusty starbursts.

\item The CO SLED of SDP.9 indicates that its molecular gas is likely dominated by one warm and dense component. The molecular gas traced by CO emission in SDP.9 is more excited than the average SMG and other lensed IR-bright sources at $z \sim 2.5$ and compatible with the excitation found in some local AGNs.

\item The total IR and dense molecular line luminosities of both SDP.9 and SDP.11 are compatible with the $L_{\rm IR} - L_{\rm dense}$ relations found for star-forming galaxies in the local Universe. This might indicate that the higher SFR in high-redshift dusty starburst is driven by a higher dense molecular gas mass fraction. Actually, the $L^{'}_{\rm HCN(1-0)}/L^{'}_{\rm CO(1-0)}$ ratio (which is significantly higher than the average ratio found for (U)LIRGs in the local Universe) suggest that a large fraction of the molecular gas in SDP.9 is in the form of dense gas.

\item The HCN/HNC ratios in SDP.9 and SDP.11 suggest that these galaxies are dominated by photon-dominated regions, similar to what it happens to most (U)LIRGs in the local Universe. Similarly, the HCN/HCO$^+$ ratios in both sources are similar to those found for local (U)LIRGs, suggesting that the dense molecular ISM in local and some high-redshift starbursts might be very similar.

\end{enumerate}

\section{Acknowledgments}

I.O., R.J.I., Z-Y.Z., and L.D. acknowledge support from the European Research Council in the form of the Advanced Investigator Programme, 321302, {\sc cosmicism}. DR acknowledges support from the National Science Foundation under grant number AST-1614213 to Cornell University. H.D. acknowledges financial support from the Spanish Ministry of Economy and Competitiveness (MINECO) under the 2014 Ram\'on y Cajal program MINECO RYC-2014-15686. IO acknowledges the warm welcome of Institute of Astrophysics of Paris (IAP), where a significant part of the analysis presented in this paper was carried out. IO also acknowledges the PNCG (Programme National Cosmologie Galaxies) for support in his stay at IAP. IO also acknowledges fruitful conversations with Padelis P. Papadopoulos and George Privon. C.F. acknowledges funding from CAPES (proc. 12203-1). MN acknowledges financial support from the European Union's Horizon 2020 research and innovation programme under the Marie Sk{\l}odowska-Curie grant agreement No 707601. This work is based on observations carried out with the VLA. The NRAO is a facility of the NSF operated under cooperative agreement by Associated Universities, Inc. Based on observations carried out with the IRAM PdBI Interferometer. IRAM is supported by INSU/CNRS (France), MPG (Germany) and IGN (Spain). This paper makes use of the following ALMA data: ADS/JAO.ALMA\#2012.1.00915.S. ALMA is a partnership of ESO (representing its member states), NSF (USA) and NINS (Japan), together with NRC (Canada) and NSC and ASIAA (Taiwan) and KASI (Republic of Korea), in cooperation with the Republic of Chile. The Joint ALMA Observatory is operated by ESO, AUI/NRAO and NAOJ. 

{\it Herschel} is an ESA space observatory with science instruments provided by European-led Principal Investigator consortia and with important participation from NASA. The Herschel spacecraft was designed, built, tested, and launched under a contract to ESA managed by the Herschel/Planck Project team by an industrial consortium under the overall responsibility of the prime contractor Thales Alenia Space (Cannes), and including Astrium (Friedrichshafen) responsible for the payload module and for system testing at spacecraft level, Thales Alenia Space (Turin) responsible for the service module, and Astrium (Toulouse) responsible for the telescope, with in excess of a hundred subcontractors. The {\it Herschel}-ATLAS is a project with {\it Herschel}. The {\it H}-ATLAS Web site is http://www.h-atlas.org/. US participants in {\it H}-ATLAS acknowledge support from NASA through a contract from JPL. SPIRE has been developed by a consortium of institutes led by Cardiff Univ. (UK) and including: Univ. Lethbridge (Canada); NAOC (China); CEA, LAM (France); IFSI, Univ. Padua (Italy); IAC (Spain); Stockholm Observatory (Sweden); Imperial College London, RAL, UCL-MSSL, UKATC, Univ. Sussex (UK); and Caltech, JPL, NHSC, Univ. Colorado (USA). This development has been supported by national funding agencies: CSA (Canada); NAOC (China); CEA, CNES, CNRS (France); ASI (Italy); MCINN (Spain); SNSB (Sweden); STFC, UKSA (UK); and NASA (USA). The SMA is a joint project between the Smithsonian Astrophysical Observatory and the Academia Sinica Institute of Astronomy and Astrophysics and is funded by the Smithsonian Institution and the Academia Sinica. The authors wish to recognize and acknowledge the very significant cultural role and reverence that the summit of Mauna Kea has always had within the indigenous Hawaiian community. We are most fortunate to have the opportunity to conduct observations from this mountain.

\bibliographystyle{mn2e}

\bibliography{ioteo_biblio}

\end{document}